\documentstyle[preprint,amsfonts,amssymb,epsfig,aps]{revtex}
\tightenlines
\newcommand{\rv}{{\bf r}}

\newcommand{\rd}{r_{12}}
\newcommand{\half}{{1\over 2}}

\newcommand{\mq}{\overline{m}}

\newcommand{\mulm}{\mu_{lm}}

\newcommand\eqref[1]{Eq.~(\ref{#1})}
\newcommand\klref[1]{(\ref{#1})}
\newcommand\cref[1]{Ref.~\cite{#1}}
\newcommand\abbref[1]{Fig.~\ref{#1}}
\newcommand\eqpref[1]{Eq.~(\protect\ref{#1})}

\newcommand\pref[1]{\protect\ref{#1}}

\newcommand{\nn}{\nonumber}

\newcommand\rhos{\rho^\ast}

\begin{document}
\draft
\title{Fluids of rod-like particles near curved surfaces}
\author{B. Groh$^1$ and S. Dietrich$^2$}
\address{
 $^1$ FOM Institute for Atomic and Molecular Physics, Kruislaan 407, 1098
 SJ Amsterdam, The Netherlands\\
 $^2$ Fachbereich Physik, Bergische Universit\"at Wuppertal, D--42097
 Wuppertal, Germany}
\date{\today}

\maketitle

%%%%%%%%%%%%%%%%%%%%%%%%%%%%%%%%%%%%%%%%%%%%%%%%%%%%%%%%%%%%%%%%%%%%%%%%%
\begin{abstract}

We study fluids of hard rods in the vicinity of hard spherical and
cylindrical surfaces at densities below the isotropic-nematic
transition. The Onsager second virial approximation is applied, which
is known to yield exact results for the bulk properties in the limit of infinitely thin
rods. This approach requires the computation of the one-particle
distribution function and of the Mayer function which is greatly
facilitated by an appropriate expansion in terms of spherical
harmonics.  We determine density
and orientational profiles as well as the surface tension $\gamma$ as
function of the surface curvature radius $R$. Already in the
low-density limit of non-interacting rods $\gamma(R)$ turns out to be
non-analytic at $1/R=0$, which prohibits the application of the commonly used
Helfrich expansion. The interparticle interaction modifies the
 behavior of $\gamma(R)$ as compared to the low-density
limit quantitatively and qualitatively.

\end{abstract}

\bigskip

%%%%%%%%%%%%%%%%%%%%%%%%%%%%%%%%%%%%%%%%%%%%%%%%%%%%%%%%%%%%%%%%%%%%%%
%%%%%%%%%%%%%%%%%%%%%%%%%%%%%%%%%%%%%%%%%%%%%%%%%%%%%%%%%%%%%%%%%%%
\section{Introduction}

A fluid of hard rods can be considered as the simplest model for
nematic liquid crystals consisting of elongated molecules. In a seminal paper in 1949
Onsager showed \cite{Onsager:49} that the steric
hard-body interactions alone can bring about an
isotropic-nematic transition. Although the steric interactions already
capture many of the essential features of liquid crystals their actual
behavior  is complicated by the presence of dispersion forces,
flexibility, dipole moments, etc. But for certain colloidal systems of
rod-like particles of synthetic or biological origin dissolved in a
suitable solvent the hard rod
model provides a quantitatively reliable effective description
\cite{Vroege:92}. Among them the ones, which are studied in most
detail, are the tabac mosaic virus and the fd-virus
with length ($L$) to diameter ($D$) ratios $L/D$ of about 17 and 150, respectively
\cite{Vroege:92}. From a theoretical point of view the limit of
infinitely thin hard rods is especially interesting because it
represents one of the very few cases for which the exact density
functional is known \cite{Onsager:49}.

Even more than in simple liquids, which are composed of spherically symmetric
particles, surface
effects are of great importance for liquid crystals. In the absence of
external fields the orientation
of the bulk fluid is  determined by its interaction with the
container walls; this phenomenon is called anchoring \cite{Jerome:91}. For
the simplest case of a hard rod fluid near a planar hard wall, as
studied theoretically
by Ho\l{}yst and Poniewierski
\cite{Holyst:88:b,Holyst:88:c,Holyst:89}, the wall induces parallel
alignment of the nematic director. An isotropic-nematic interface also
aligns the nematic phase parallel to the interface for large aspect
ratios $L/D$ while a non-trivial tilt angle arises for lower aspect
ratios \cite{Moore:90}. In view of the substantial technical
difficulties which are associated with the theoretical description on
a truly microscopic scale, especially for curved surfaces, it is a
natural first step to analyze the interface between the {\it isotropic}
phase and a hard wall. The corresponding density
and orientational order profiles near a {\it planar} wall have been determined by Poniewierski
\cite{Poni:93} in the framework of the Onsager theory, who also found
indications for spontaneous biaxial order at the surface already below
the bulk transition to the nematic phase. Mao et al. have compared this theory with computer
simulations for finite aspect ratios $L/D$
\cite{Mao:97:a} and have calculated the depletion force between planar
walls or large spheres immersed in a solution of rods
\cite{Mao:95,Mao:97}, using the Derjaguin approximation. 

In the
present work we focus on the orientational and positional order as
well as the surface tension near {\it curved} hard walls, taking into
account the steric interactions between the rods. The
curvature has an appreciable effect on the structure and the
thermodynamics of the fluid  if the radius of curvature $R$ is of
the order of the particle length $L$. Accordingly as possible applications
one can think of the following systems: (i) rod-like particles
confined to the interior of small pores within porous materials; (ii) colloidal
suspensions of rods that contain a second, diluted, component of
larger, e.g., spherical particles; (iii) membranes, especially
vesicles, immersed in colloidal rod solutions, resembling, e.g.,
solutions of viruses. The curvature
dependent surface tension and the depletion forces in case (ii) have
been determined by Auvray \cite{Auvray:81} and Yaman et
al. \cite{Yaman:97:PRL,Yaman:97} for fluids of non-interacting rods  corresponding to the limit of infinite dilution. Their most
surprising result was that the surface free energy does not contain a
term linear in the curvature $1/R$ and that the quadratic term has different
amplitudes for different signs of the curvature. This non-analyticity
prohibits the application of the common Helfrich expansion. Thus for case (iii) above the
effect of the rods on the elastic properties of the membranes cannot be described by a renormalization of the
bending rigidities as it is possible for membranes exposed to a suspension
of  spherical colloidal particles \cite{Yaman:97:PRL} or
polymers \cite{Eisenriegler:96}. In order to be able to assess the range of
validity of the results obtained in the ideal limit of non-interacting rods, in the present
paper we tackle the full problem including the inter-particle interactions by
employing the Onsager density-functional theory (Sec.~\ref{DFT}),
which yields density profiles (Sec.~\ref{ELG}) and the surface tension
(Sec.~\ref{surf}). Our main results are summarized in
Sec.~\ref{summary} while technical details are presented in Appendices
A and B.

%%%%%%%%%%%%%%%%%%%%%%%%%%%%%%%%%%%%%%%%%%%%%%%%%%%%%%%%%%%%%%%%%%%
\section{Model and density-functional theory} \label{DFT}

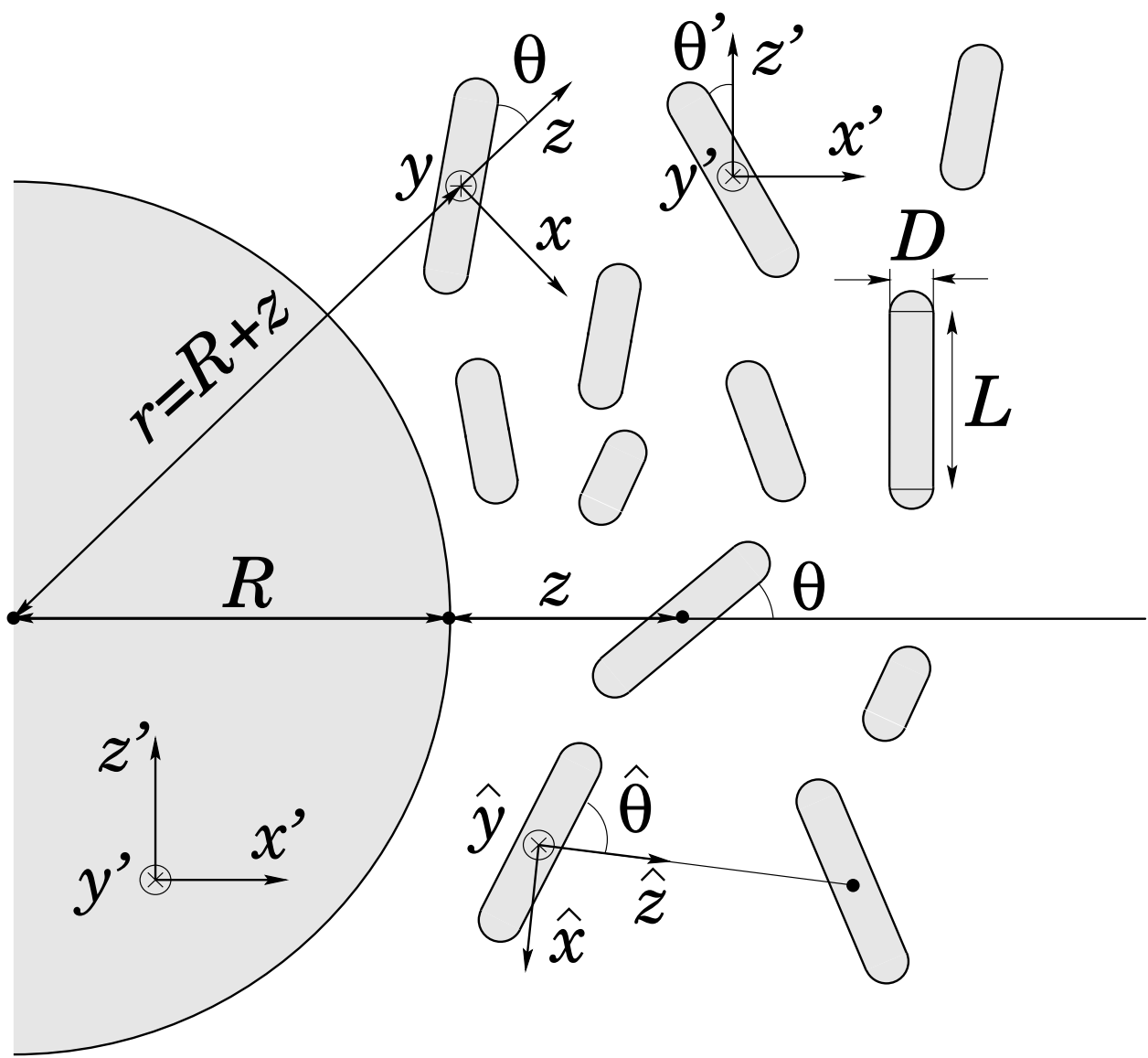
\begin{figure}
\caption{The  system under consideration consists of a fluid of
monodisperse hard
spherocylinders of diameter $D$ and length $L$ in contact with a
spherical or cylindrical hard wall of radius $R$. We assume that the
orientational distribution only depends on the normal distance $z$
and, for a spherical wall,
the angle $\theta$ of the particle axis with respect to the surface
normal. During the evaluation of the excess free energy the particle
orientations are described in three different reference frames: the
frame $x'y'z'$ fixed in space, the frame $xyz$ fixed by the local normal
direction, and the interparticle frame $\hat x\hat y\hat z$. The
rotation between the latter two is described by the Euler angles
$\psi_i$, $\eta_i$, and $\chi_i$ (see \eqpref{Ylmtrafo}). The
different sizes of the spherocylinders indicate that the particles
typically do not lie in the $x'y'$ plane and thus only their
projection onto that plane is shown. For reasons of clarity only the
polar angles $\theta'$, $\theta$, and $\hat\theta$ are shown, but not
the corresponding azimuthal angles $\phi'$, $\phi$, and $\hat\phi$.}
\label{fig:system}
\end{figure}

Based on density-functional theory we study a fluid of hard spherocylinders of length $L$ and diameter
$D$ in the vicinity of a hard spherical or cylindrical surface of
radius $R$ (see \abbref{fig:system}). In order to keep the numerical difficulties
tractable we restrict ourselves to the limit $D/L\to 0$ with
$R/L$ fixed. The number density of the centers of mass of these thin rods at a
point $\rv$ with orientation $\omega'=(\theta',\phi')$ is denoted by
$\hat\rho(\rv,\omega')$. The corresponding grand canonical functional is given by 
\begin{eqnarray} \label{Omges}
  \beta\Omega[\{\hat\rho(\rv,\omega')\}]&=&\int d^3r d\omega'
  \hat\rho(\rv,\omega') \left[\ln \left(4\pi \lambda^3 \hat\rho(\rv,\omega')\right)-1
  -\beta\mu+\beta V(\rv,\omega')\right] \nn \\
 & &{}+\beta
  F_{ex}[\{\hat\rho(\rv,\omega')\}].
\end{eqnarray}
Here $k_B \beta=1/T$ is the inverse temperature, $\mu$ the chemical
potential, $V(\rv,\omega')$ the external potential exerted by the hard
wall and $\lambda$ the thermal de Broglie wavelength. Within the
Onsager second virial approximation the excess free energy $F_{ex}$ is
\cite{Onsager:49}
\begin{equation} \label{Fex}
 \beta F_{ex}[\{\hat\rho(\rv,\omega')\}]=-\half \int d^3r_1 d\omega'_1
 d^3r_2 d\omega'_2\, \hat\rho(\rv_1,\omega'_1) \hat\rho(\rv_2,\omega'_2)
 f(\rv_{12},\omega'_1,\omega'_2)
\end{equation}
with the interparticle vector $\rv_{12}=\rv_2-\rv_1$ and the Mayer
function $f(\rv_{12},\omega'_1,\omega'_2)$ which for hard particles equals $-1$ if the two
particles overlap and zero otherwise. Onsager demonstrated that this
approximation becomes {\it exact} in the limit $D/L\to 0$ for the bulk
properties
\cite{Onsager:49}, and this is expected to hold also for surface quantities
\cite{Poni:93,Mao:97}. For a planar surface this approach has turned
out to be
quantitatively reliable for $D/L\lesssim 0.1$ \cite{Mao:97}.

As mentioned in the introduction, the corresponding problem with non-interacting particles, i.e., in the
presence of the hard wall interaction but with
$F_{ex}=0$, has been analyzed by Yaman et al. \cite{Yaman:97:PRL,Yaman:97}. Taking
into account the inter-particle interaction increases the complexity
considerably due to the ensuing non-locality and the high dimensional
integration in \eqref{Fex}. In the remaining part of this section we
evaluate the expression for $F_{ex}$ by exploiting the symmetries of the
density profile $\hat\rho(\rv,\omega)$ using appropriate expansions in
terms of
spherical harmonics. In Sec.~\ref{ELG} the thermodynamically stable
equilibrium profile is  obtained  by minimization
of the density functional, which amounts to solving an
integral equation. The value of the functional at this minimum is the
grand canonical potential of the inhomogeneous fluid from which the
surface tension at the curved walls is determined in Sec.~\ref{surf}.

The particle orientation is conveniently described in a local reference
frame whose polar axis is that wall normal which runs through the rod
center and whose $y$ axis, in the
case of a cylindrical wall, is aligned with the cylinder axis (see \abbref{fig:system}).
Since we are interested in bulk densities $\rho_b$ below the
isotropic-nematic transition, i.e., $\rho_b<4.2 (DL^2)^{-1}$ \cite{Onsager:49}, we may assume that the number density
$\rho(r)=\int d\omega' \hat\rho(\rv,\omega')$ as well as the
orientational distribution measured in the local reference system
(denoted by $\omega=(\theta,\phi)$) depend only on the radial coordinate $r$ in a
spherical or cylindrical coordinate system, which allows us to make the
following ansatz
\begin{equation} \label{rhoS}
  \hat\rho^S(\rv,\omega)=\frac{\rho_b}{2\pi}\sum_{l=0}^\infty
  \alpha_l(r) P_l(\cos\theta)
\end{equation}
and
\begin{equation} \label{rhoC}
  \hat\rho^C(\rv,\omega)=\rho_b \sum_{l=0}^\infty \sum_{m=-l}^l
  \mu_{lm}(r) Y_{lm}(\omega)
\end{equation}
for a sphere (S) and a cylinder (C), respectively. The functions $P_l$
and $Y_{lm}$ are Legendre polynomials and spherical harmonics, respectively. For
large distances from the wall the fluid is isotropic
($\hat\rho(\rv,\omega)=\rho_b/4\pi$), so that
$\alpha_l(r\to\infty)=\half \delta_{l,0}$ and
$\mu_{lm}(r\to\infty)=(4\pi)^{-1/2} \delta_{lm,00}$. At a sphere the
density does not depend on the azimuthal angle $\phi$. At a
cylinder  the symmetries
\begin{equation}
  \hat\rho(r,\theta,\phi)=\hat\rho(r,\pi-\theta,\phi)=\hat\rho(r,\theta,-\phi)=\hat\rho(r,\theta,\pi-\phi)
\end{equation}
imply that  $\mu_{lm}=0$ if $l$ or $m$ is odd and
$\mu_{lm}=\mu_{l\mq}=\mulm^\ast$ for $l$ and $m$ even (here and in the
following $\mq=-m$).

The coordinates $\omega$ can be expressed in terms of  the  coordinates
$\omega'$ corresponding to a frame fixed in space by a (position dependent) rotation. Therefore the angular
integrations in Eqs.~\klref{Omges} and \klref{Fex} can  be
taken also over $\omega$. However, the Mayer function is naturally
expressed within a third coordinate system $\hat\omega$ given by
the interparticle vector $\rv_{12}$. In order to perform the angular
integrations  in the following we will determine the
transformation from $\hat\omega$ to $\omega$. The
definition of the different reference frames is illustrated in
\abbref{fig:system}.

For any uniaxial molecule the Mayer function can be expanded as
\cite{Gray}
\begin{equation} \label{fser}
  f(\rd,\hat\omega_1,\hat\omega_2)=\sum_{l_1,l_2,m} f_{l_1l_2m}(\rd)
  Y_{l_1m}(\hat\omega_1) Y_{l_2\mq}(\hat\omega_2).
\end{equation}
The solid angles $\hat\omega_i$ refer to a particle fixed
reference system (see \abbref{fig:system}) with its $z$ axis parallel to the interparticle
vector $\rv_{12}$ (and arbitrary $x$ axis). The determination of the
expansion coefficients 
\begin{equation} \label{fllmdef}
  f_{l_1l_2m}(\rd)=\int d\hat\omega_1 d\hat\omega_2
  f(\rd,\omega_{12}=0,\hat\omega_1,\hat\omega_2)
  Y_{l_1m}^\ast(\hat\omega_1)  Y_{l_2\mq}^\ast(\hat\omega_2)
\end{equation}
is discussed in Appendix~\ref{fllm}.
Spherical harmonics in different reference systems are related via the
rotation matrices $D_{nm}^l$ \cite{Gray}:
\begin{equation} \label{Ylmtrafo}
  Y_{lm}(\omega_i)=\sum_n D_{nm}^l(\psi_i,\eta_i,\chi_i)
  Y_{ln}(\hat\omega_i), \qquad i=1,2,
\end{equation}
where the Euler angles $\psi_i(\rv_1,\rv_2)$, $\eta_i(\rv_1,\rv_2)$,
and $\chi_i(\rv_1,\rv_2)$ describe
the rotation of the particle based axes ($\hat\omega_i$) onto the
surface normal based axes ($\omega_i$) (for the definition of the
Euler angles see, e.g., Fig.~A.6 in \cref{Gray}).  Inserting
Eqs.~\klref{rhoS}, \klref{fser}, and \klref{Ylmtrafo} into \eqref{Fex}
yields for the spherical case
\begin{eqnarray}
 \beta F_{ex}^S&=&-\frac{\rho_b^2}{2\pi} \sum_{l_1,l_2,m}
 [(2l_1+1)(2l_2+1)]^{-1/2} \\
 & &{}\times \int d^3r_1 d^3r_2\, \alpha_{l_1}(r_1)
 \alpha_{l_2}(r_2) f_{l_1l_2m}(\rd)
 D_{m0}^{l_1\ast}(\psi_1,\eta_1,\chi_1)
 D_{\mq0}^{l_2\ast}(\psi_2,\eta_2,\chi_2) \nn.
\end{eqnarray}
One still has the freedom to fix the orientation of the $y$ axes in the
different reference systems. If one chooses them to be all parallel to
each other and perpendicular to the plane spanned by $\rv_1$ and
$\rv_2$ the transformations described by the Euler angles become
simple rotations around the $y$ axis so that $\psi_i=\chi_i=0$. With
\cite{Gray}
\begin{equation}
  D_{m0}^{l\ast}(0,\eta,0)=\sqrt{\frac{4\pi}{2l+1}} Y_{lm}(\eta,0)
\end{equation}
 one finds for a system of radial size $\cal L$,
outside of a spherical cavity of radius $R$
\begin{eqnarray}
  \beta F_{ex}^S&=&-(4\pi\rho_b)^2 \sum_{l_1,l_2,m}
 [(2l_1+1)(2l_2+1)]^{-1} \int_R^{R+{\cal L}} dr_1\,r_1^2\int_R^{R+{\cal L}} dr_2\,
 r_2^2 \alpha_{l_1}(r_1) \alpha_{l_2}(r_2) \\
 & &{}\times \int_{-1}^1 d\cos\gamma\,
 f_{l_1l_2m}(\rd) Y_{l_1m}(\eta_1,0) Y_{l_2\mq}(\eta_2,0) \nn.
\end{eqnarray}
The angles $\gamma$, $\eta_1$, and $\eta_2$ are those between the
vectors $\rv_1$ and $\rv_2$, $\rv_1$ and $\rv_{12}$, and $\rv_2$ and
$\rv_{12}$, respectively, and
$\rd=(r_1^2+r_2^2-2r_1r_2\cos\gamma)^{1/2}$. If we use $\rd$ instead
of $\cos\gamma$ as integration variable we finally obtain 
\begin{equation} \label{FexS}
  \beta F_{ex}^S=\half \rho_b^2 \sum_{l_1,l_2} \int_R^{R+{\cal L}} dr_1\,r_1
 \int_R^{R+{\cal L}} dr_2\,r_2 \alpha_{l_1}(r_1) \alpha_{l_2}(r_2)
 w_{l_1l_2}(r_1,r_2)
\end{equation}
with
\begin{equation} \label{wS}
  w_{l_1,l_2}(r_1,r_2)=-\frac{32\pi^2}{(2l_1+1)(2l_2+1)} \sum_m
  \int_{|r_1-r_2|}^{r_1+r_2} d\rd \rd f_{l_1l_2m}(\rd)
  Y_{l_1m}(\theta_1,0) Y_{l_2\mq}(\theta_2,0)
\end{equation}
and
\begin{equation}
  \cos\eta_1=\frac{r_2^2-r_1^2-\rd^2}{2 r_1 \rd}, \qquad
  \cos\eta_2=\frac{r_2^2-r_1^2+\rd^2}{2 r_2 \rd}.
\end{equation}

An equivalent expression for a {\it planar} wall with surface area $A$
can be derived along the
same lines. In this case the Euler angles are the same for both
particles because the direction of the surface normal is the same
everywhere. Again one can choose $\psi_i=\chi_i=0$ and finds
\begin{equation} \label{FexP}
  \beta F_{ex}^P/A=\half \rho_b^2 \sum_{l_1,l_2} \int_0^{\cal L} dz_1
  \int_0^{\cal L} dz_2\, \alpha_{l_1}(z_1) \alpha_{l_2}(z_2)
  w_{l_1l_2}^P(z_1-z_2).
\end{equation}
where the interaction kernel $w_{l_1l_2}^P$ now depends only on one
variable:
\begin{equation} \label{wP}
  w_{l_1l_2}^P(z_{12})=-\frac{8\pi}{(2l_1+1)(2l_2+1)} \sum_m
  \int_{|z_{12}|}^\infty d\rd\, \rd f_{l_1l_2m}(\rd)
  Y_{l_1m}(\eta,0)
  Y_{l_2\mq}(\eta,0).
\end{equation}
with $\eta=\arccos(z_{12}/\rd)$.
It can be shown that $w_{l_1,l_2}(R+z_1,R+z_2)=4\pi
w_{l_1l_2}^P(z_1-z_2)+O(1/R)$ for $z_1,z_2\ll R$.

The cylindrical case is considerably complicated by the lower symmetry
of $\hat\rho(r,\omega)$. Using Eqs.~\klref{Fex}, \klref{rhoC}, and
\klref{Ylmtrafo} and performing the integrations over $\omega_1$,
$\omega_2$, as well as $z_1$ and $\tilde\phi_1$, where
$\rv_i=(r_i,z_i,\tilde\phi_i)$, $i=1,2$, in cylindrical coordinates, one obtains
\begin{equation} \label{FexC}
  \beta F_{ex}^C/H=\half\rho_b^2 \int_R^{R+{\cal L}} dr_1 r_1 \int_R^{R+{\cal L}}
  dr_2 r_2 \sum_{l_1,l_2,m_1,m_2} \mu_{l_1m_1}(r_1) \mu_{l_2m_2}(r_2)
  w_{l_1m_1l_2m_2}(r_1,r_2)
\end{equation}
with
\begin{equation}
  w_{l_1m_1l_2m_2}(r_1,r_2)=-2\pi \sum_m \int_{-\infty}^\infty dz_{12}
  \int_0^{2\pi} d\tilde\phi_{12} f_{l_1l_2m}(\rd)
  D_{mm_1}^{l_1\ast}(\psi_1,\eta_1,\chi_1) D_{\mq
  m_2}^{l_2\ast}(\psi_2,\eta_2,\chi_2).
\end{equation}
Here $H$ is the macroscopic height of the cylinder and
$z_{12}=z_2-z_1$, $\tilde\phi_{12}=\tilde\phi_2-\tilde\phi_1$,
$\rd=(r_1^2+r_2^2+z_{12}^2-2 r_1 r_2 \cos\tilde\phi_{12})^{1/2}$.
In a rather lengthy calculation the dependence of the Euler angles on $r_1$, $r_2$, $z_{12}$, and
$\tilde\phi_{12}$ can be worked out by decomposing the rotations that
connect the different reference systems into three successive simple
rotations around (intermediate) coordinate axes (see App.~A.2 in
\cref{Gray}). It is helpful to use $\rd$ and $u=\rd^2-z_{12}^2$ as the
integration variables which leads to
\begin{eqnarray} \label{wC}
  w_{l_1m_1l_2m_2}(r_1,r_2)&=&-16\pi \sum_m \int_{|r_1-r_2|}^\infty d\rd
  \rd f_{l_1l_2m}(\rd)  \nn\\
  & &{}\times \int_{(r_1-r_2)^2}^{\rd^2} du \left[
  (\rd^2-u)((r_1+r_2)^2-u)(u-(r_1-r_2)^2)\right]^{-1/2} \\
  & &{}\times
  \cos\left[ m(\psi_1-\psi_2)+m_1\chi_1+m_2\chi_2\right]
  d_{mm_1}^{l_1}(\eta_1) d_{\mq m_2}^{l_2}(\eta_2). \nn
\end{eqnarray}
Here the rotation
matrices have been written as \cite{Gray}
\begin{equation}
  D_{mn}^l(\psi,\eta,\chi)=e^{-im\psi} d^l_{mn}(\eta) e^{-in\chi}
\end{equation}
where the functions $d^l_{mn}$ can be calculated by using Eq.~(A.65) in
\cref{Gray}.
The advantage of \eqref{wC} is that the inner integral can be evaluated
without the time-consuming calculation of $f_{l_1l_2m}$. In these variables
the Euler angles are
\begin{eqnarray} 
  \cos\eta_1=\frac{r_2^2-r_1^2-u}{2\rd r_1} &\qquad &
   \cos\eta_2=\frac{r_2^2-r_1^2+u}{2\rd r_2}  \\
  \tan\psi_1=2 r_1 \Delta\cos\eta_1  &\qquad &
  \tan\psi_2=2 r_2 \Delta\cos\eta_2   \\
  \tan\chi_1=-2 r_1 \Delta &\qquad &
  \tan\chi_2=-2 r_2 \Delta 
\end{eqnarray}
with
\begin{equation}
  \Delta=\left(\frac{\rd^2-u}{((r_1+r_2)^2-u)(u-(r_1-r_2)^2)}\right)^{1/2}.
\end{equation}

For the inside of a sphere (cylinder) the integration range for the
radial integrals  in \eqref{FexS} (\eqref{FexC}) has to be replaced
by $[0,R]$.

%%%%%%%%%%%%%%%%%%%%%%%%%%%%%%%%%%%%%%%%%%%%%%%%%%%%%%%%%%%%%%%%%%%%%%%%%%%%
%%%%%%%%%%%%%%%%%%%%%%%%%%%%%%%%%%%%%%%%%%%%%%%%%%%%%%%%%%%%%%%%%%%%%%%%%%%%
\section{Density profiles} \label{ELG}

The equilibrium density profile minimizes the grand-canonical
functional, i.e., it is a solution of
$\frac{\delta\Omega}{\delta\hat\rho(r,\omega)}=0$ under the boundary
condition $\hat\rho(r,\omega)\to \rho_b/4\pi$ for $r\to\infty$. By
using the relation
\begin{equation}
  \frac{\delta
  \alpha_l(r')}{\delta\hat\rho(r,\omega)}=\frac{2l+1}{2\rho_b}
  \delta(r-r')  P_l(\cos\theta)
\end{equation}
and exploiting the symmetry property $w_{l_1l_2}(r_1,r_2)=w_{l_2l_1}(r_2,r_1)$ one finds
for the spherical wall the Euler Lagrange equation 
\begin{equation} \label{ELGbare}
 4\pi\lambda^3 \hat\rho(r,\theta)=\exp\left[\beta\mu-\beta V(r,\theta)
 -\frac{\rho_b}{4\pi r} \sum_{l_1,l_2} \frac{2 l_1+1}{2}
 P_{l_1}(\cos\theta) \int dr' r' \alpha_{l_2}(r') w_{l_1l_2}(r,r') \right].
\end{equation}
With $\hat\rho_0(r,\theta)=(4\pi\lambda^3)^{-1}
\exp[\beta\mu-V(r,\theta)]$ as the corresponding profile for
non-interacting rods at the same chemical potential one finds that
\begin{equation} \label{tilderho}
  \tilde\rho(r,\theta):=\hat\rho(r,\theta)/\hat\rho_0(r,\theta)
  =:\sum_l \beta_l(r) P_l(\cos\theta)
\end{equation}
satisfies
\begin{equation} \label{eqtrho}
  \tilde\rho(r,\theta)=\exp\left[-\sum_l P_l(\cos\theta)
  p_l(r)\right]
\end{equation}
with
\begin{equation} \label{pl}
  p_l(r)=\frac{\rho_b}{4\pi r} \frac{2l+1}{2} \sum_{l'} \int dr'\, r'
  \alpha_{l'}(r') w_{ll'}(r,r').
\end{equation}
(Strictly speaking $\tilde\rho$ cannot be defined by \eqref{tilderho}
for the forbidden orientations, for which both $\hat\rho_0$ and
$\hat\rho$ vanish. Instead we {\it define} it by Eqs.~\klref{eqtrho} and
\klref{pl} in this region.)
The function $\hat\rho_0(r,\theta)$ equals
$\rho_0/4\pi$ for orientations that are allowed by the hard wall and
zero otherwise. The density $\rho_0$ corresponding to the chemical
potential $\mu$ follows from the bulk limit of the density functional.
For an isotropic fluid in a volume $V$ one has
\begin{equation} \label{omb}
  \beta\Omega/V=\beta\omega_b=\rho_b
  \left(\ln\lambda^3\rho_b-1-\beta\mu+\half\rho_b^2 v_0\right)
\end{equation}
with 
\begin{equation} \label{v0}
v_0=-\frac{1}{(4\pi)^2} \int d^3\rd d\omega_1 d\omega_2\,
f(\rv_{12},\omega_1,\omega_2)=\frac{\pi}{2} D L^2.
\end{equation}
The same equation
without the last term holds for the ideal gas limit. Minimization yields $\rho_0=\rho_b \exp(\rho_b v_0)$. The allowed values of
$\theta$ for given $r$ and $R$ are determined in
Appendix~\ref{rhoid}. Thus based on the known function
$\hat\rho_0(r,x=\cos\theta)$ the coefficients $\alpha_l$ in \eqref{pl}
can be expressed in terms of the coefficients $\beta_l$ introduced in \eqref{tilderho}:
\begin{equation} \label{a2b}
  \alpha_l(r)=\frac{2l+1}{2} \frac{2\pi}{\rho_b} \sum_{l'}
  \beta_{l'}(r) \int_{-1}^1 dx P_l(x) P_{l'}(x) \hat\rho_0(r,x)
\end{equation}
where the integration over $x$  can be carried out analytically for given $l$
and $l'$. This allows one to calculate the coefficients $\beta_l$ by
solving iteratively the following system of equations together with Eqs.~\klref{pl} and \klref{a2b}: 
\begin{equation} \label{ELGbeta}
  \beta_l(r)=\frac{2l+1}{2} \int dx P_l(x) \exp\left[-\sum_{l'}
  P_{l'}(x) p_{l'}(r)\right].
\end{equation}
The advantage of first seeking the solution for $\tilde\rho$ instead
of $\hat\rho$ is that the former function is smoother near the
transition from allowed to forbidden orientations and hence can be
better approximated with a limited number of Legendre polynomials.

The cylindrical case can be treated completely analogously. With the
expansion $\tilde\rho(r,\omega)=\sum_{lm} \nu_{lm}(r) Y_{lm}(\omega)$
one obtains
\begin{equation}
  \nu_{lm}(r)=\int d\omega Y_{lm}^\ast(\omega) \exp\left[-\sum_{l',m'}
  Y_{l'm'}^\ast(\omega) p_{l'm'}(r)\right]
\end{equation}
with
\begin{equation}
  p_{lm}(r)=\frac{\rho_b}{2\pi} \sum_{l',m'} \int dr' r'
  \mu_{l'm'}(r') w_{lml'm'}(r,r')
\end{equation}
and
\begin{equation}
  \mu_{lm}(r)=\frac{1}{\rho_b}\sum_{l',m'} \nu_{l'm'}(r) \int d\omega
  Y_{lm}^\ast(\omega) Y_{l'm'}(\omega) \hat\rho_0(r,\omega).
\end{equation}
These equations are valid both for the outside and the inside of the sphere
or cylinder if the $r'$ integrations are taken over the interval $[R,\infty)$ or
$[0,R]$, respectively. But we note that the functions $\hat\rho_0$ have
completely different forms in these two cases (see Appendix~\ref{rhoid}).
It is assumed that the fluid inside a spherical or cylindrical cavity
is in equilibrium with a particle reservoir at the chemical potential
$\mu$ corresponding to the bulk density $\rho_b$, which is kept fixed
when $R$ is varied. For small radii the actual density at the center of the cavity
may differ from $\rho_b$ although this effect is certainly numerically
neglegible in the examined range of radii $|R|\geq 3$.

In practice we have truncated all $l$ sums at $l_{max}=10$
($l_{max}=8$) for spheres (cylinders) and the radial integrals were
cut off at a distance ${\cal L}=2 L$ (${\cal L}=1.5 L$) from the
wall. Beyond this distance the profile was assumed to take on its bulk
value and corresponding asymptotic corrections were added to $p_l$
($p_{lm}$) in the vicinity of the cutoff. A step size of $\Delta
r=0.02 L$ ($\Delta r=0.03125 L$) was used for all functions of
$r$. First the values of $w_{l_1l_2}(r_1,r_2)$
($w_{l_1m_1l_2m_2}(r_1,r_2)$) were calculated and stored for all
necessary values of $r_1$ and $r_2$ and of the indices. This step
required by far the largest fraction of the computer time. Thereafter
for a series of bulk densities $\rho_b$ the coefficients $\beta_l$
($\nu_{lm}$) were determined by a simple Picard iteration scheme with
retardation.

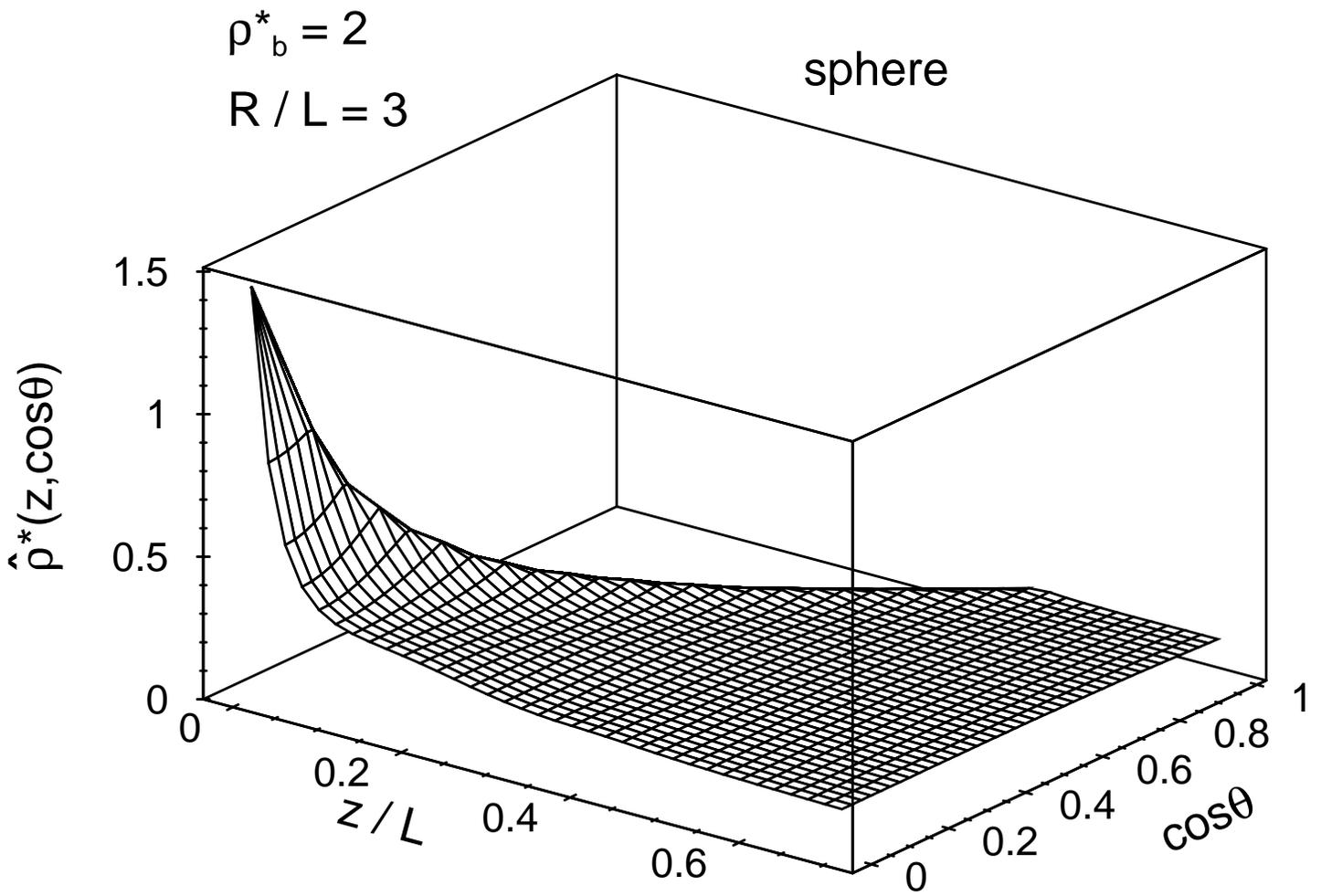
\begin{figure}
\caption{Full density profile
$\hat\rhos(z,\cos\theta)=\hat\rho(z,\cos\theta) DL^2$ 
outside a sphere of radius $R/L=3$ 
for the bulk density $\rhos_b=\rho_b D L^2=2$. At small distances $z$ from the wall
large values of $\cos\theta$ are forbidden due to overlap. Therefore
the profile is exactly zero behind this ``step''.  In the allowed
region the most prominent feature induced by the interaction between
the rods is the strong increase of the
density at small distances $z$. For fixed $z$ orientations close to the
step, i.e., those with one end of the rod touching the wall, are
favored. If the interparticle interactions were neglected the profile
would be constant in the whole accessible region.}
\label{fig:rhotld}
\end{figure}

In the following a negative (positive) radius $R$ signifies that the wall curves
towards (away from) the fluid, and $z$ is the distance from the surface. As
reduced density we employ $\rhos=\rho D L^2$; in these units the
isotropic-nematic transition takes place at $\rhos_b\simeq 4.2$
\cite{Onsager:49}, which provides an upper limit for the present
approach because in the nematic phase the orientational structure does
not exhibit the  symmetries assumed here. A typical density profile
$\hat\rho(z,\cos\theta)$ outside of a sphere is shown in
\abbref{fig:rhotld}. For $z<L/2$ orientations with large $\cos\theta$
are forbidden so that the profile has a discontinuity along the line
$\cos\theta=x_{max}(z)$ determined in Appendix~\ref{rhoid}. When the rods
do not interact among each other, i.e., for $\rho_b\to 0$, all allowed orientations
have the same probability. The presence of the steric interaction
induces a {\it strong} increase of the density close to the surface,
while there is only a weak  dependence on $\cos\theta$ within the allowed
region. Orientations near the discontinuity, where one end of the rod
touches the wall, are slightly favored. Note that no packing effects
are visible. These will occur on the much smaller length scale $D$ and
presumably only at much higher densities where the packing fraction
$\Phi\sim \rho D^2 L$ is of order unity. The profiles for other radii,
even for the opposite sign of the curvature, look essentially the same. In
the latter case there is a very small region close to the surface that
is not accessible to any rod center. For a cylindrical wall the
profiles also depend on the azimuthal angle, but except very close to
 surfaces with negative curvatures this dependence is very weak and a
plot of $\hat\rho^C(z,\cos\theta,\phi)$ for any fixed $\phi$ looks
very similar to \abbref{fig:rhotld}.

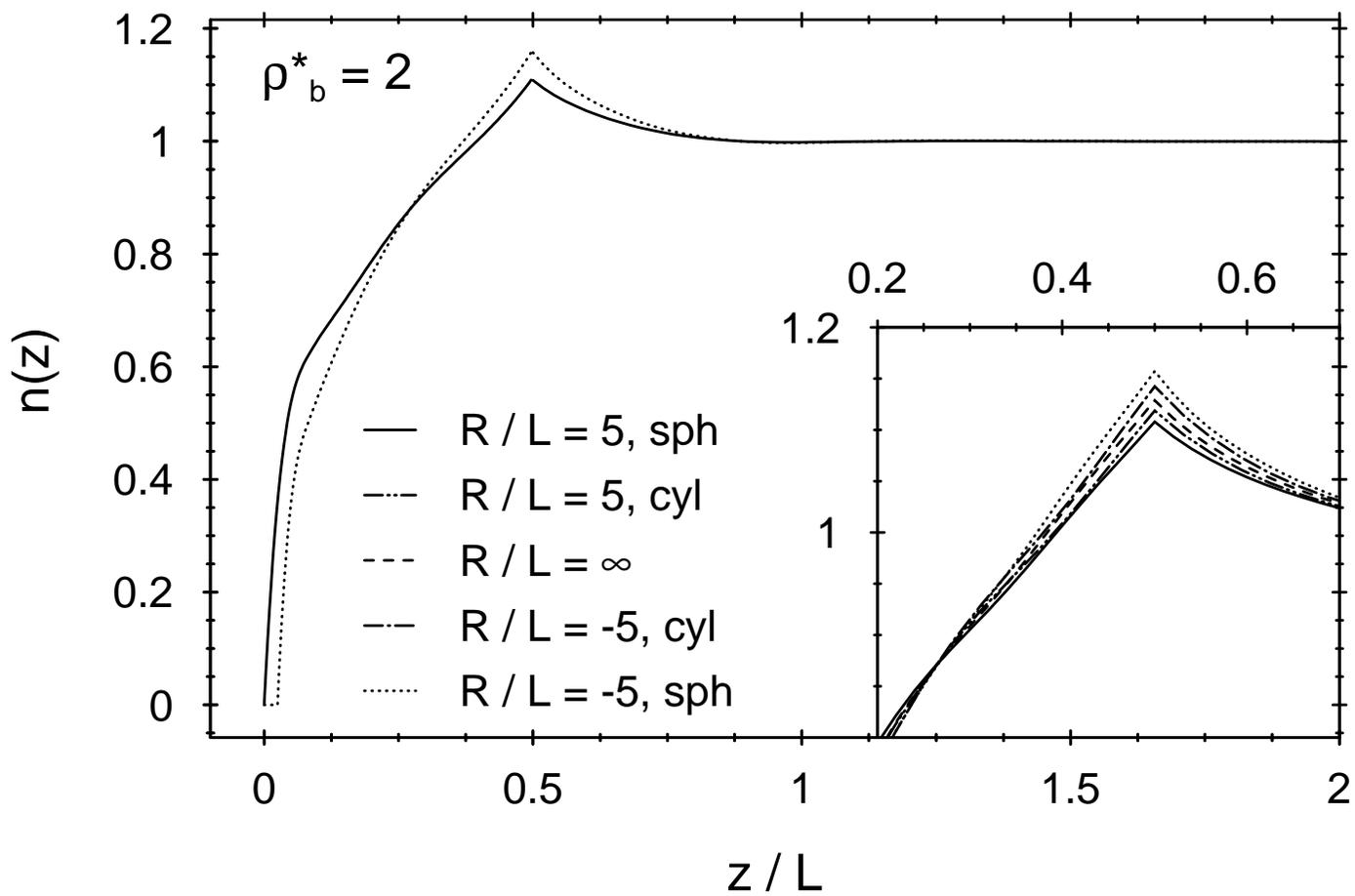
\begin{figure}
\caption{Normalized orientationally averaged density $n(z)$ for fluids
in contact with planar, cylindrical, and spherical walls of positive
and negative curvature for a fixed bulk density $\rhos_b=2$. The inset
shows the behavior in the vicinity of the cusp which occurs at $z=L/2$
and is followed by a rapid decay towards the bulk limit $n(z)=1$. All
curves have about the same value at $z/L\simeq0.27$ but they do not
intersect exactly at one point. For better visibility only two curves are
shown in the main part of the figure. The remaining profiles lie in
between these two. Close to the wall the number density is larger for
positive curvature. Inside a sphere $n(z)=0$ for very small $z$
because the centers of the rods cannot come arbitrarily close to the
wall. Small kinks in the small $z$ range that are caused by the
numerical discretization were removed by fitting a smooth curve to the
raw data.}
\label{fig:ponin}
\end{figure}

The normalized orientationally averaged number density is defined by
\begin{equation} \label{nz}
n(z)=\int d\omega \hat\rho(z,\omega)/\rho_b.
\end{equation}
This function increases
for small $z$ up to $z=L/2$ where it exhibits a cusp and then rapidly
decreases to its bulk limit 1 which is essentially reached already at
$z=L$. As shown in \abbref{fig:ponin} within the examined range of
curvatures ($|R|/L\gtrsim 3$) it depends only slightly on $R$. If $1/R$ is
decreased $n(z)$ becomes smaller for $z/L\lesssim0.27$ and larger for
$z/L\gtrsim0.27$. The results for the cylinder lie between those for the
planar wall ($R=\infty$) and for a sphere with the same radius. Due to
the finite step size $\Delta r$ and the steepness of
$\hat\rho(z,\omega)$ the raw data for $n(z)$ exhibit visible kinks at
$z=n\Delta r$ for small integers $n$. These have been removed from
\abbref{fig:ponin} by
fitting of an appropriate smooth function to the data.

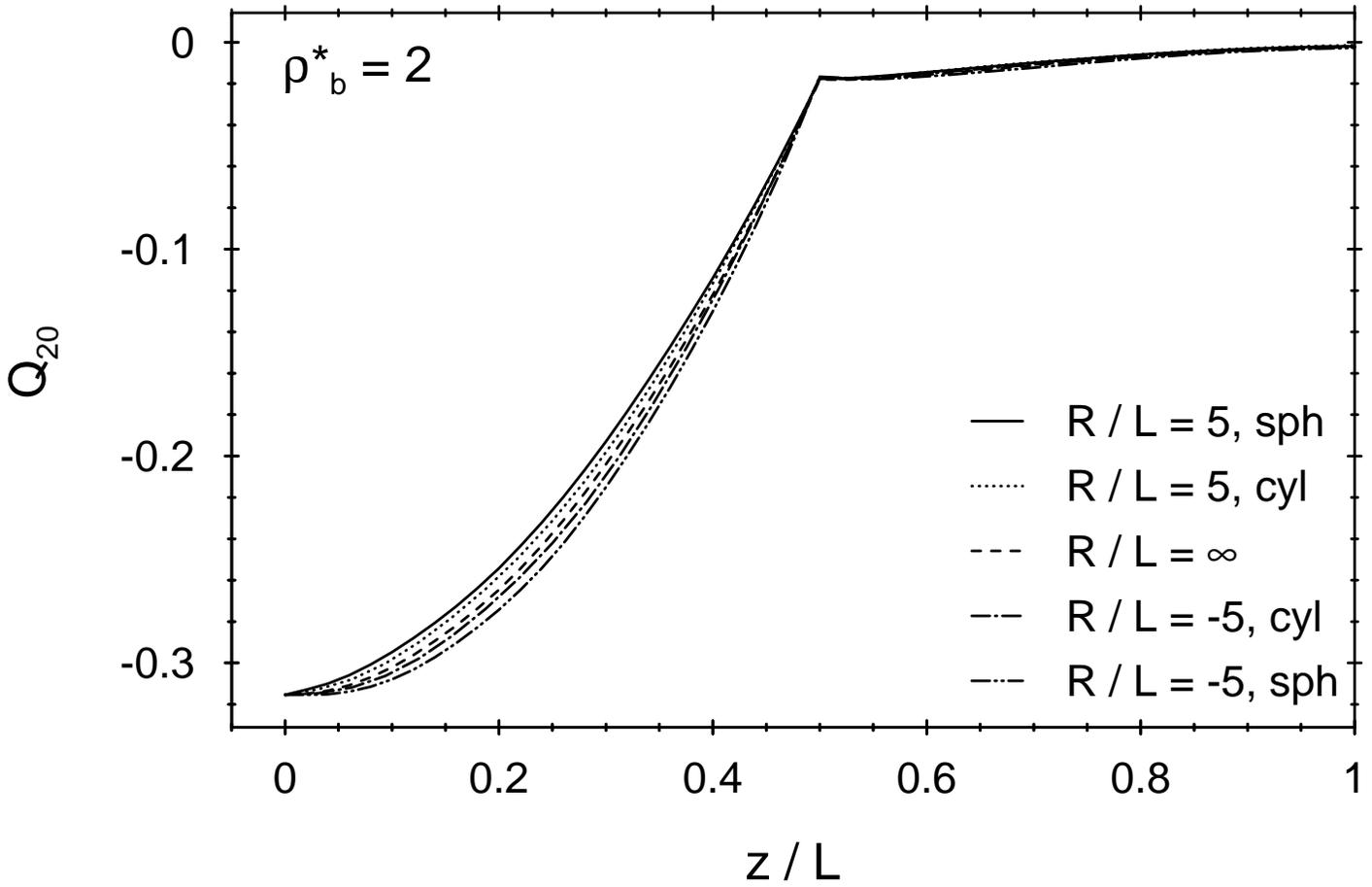
\begin{figure}
\caption{Uniaxial nematic order parameter $Q_{20}$ as defined in
\eqpref{Qlm} for a fluid outside a sphere of radius $R/L=5$ for
different densities. The tendency for parallel orientations, as indicated
by negative values of $Q_{20}$, becomes more pronounced and longer ranged
with increasing density. $Q_{20}(z=0)=-0.3154$.}
\label{fig:poniqrho}
\end{figure}

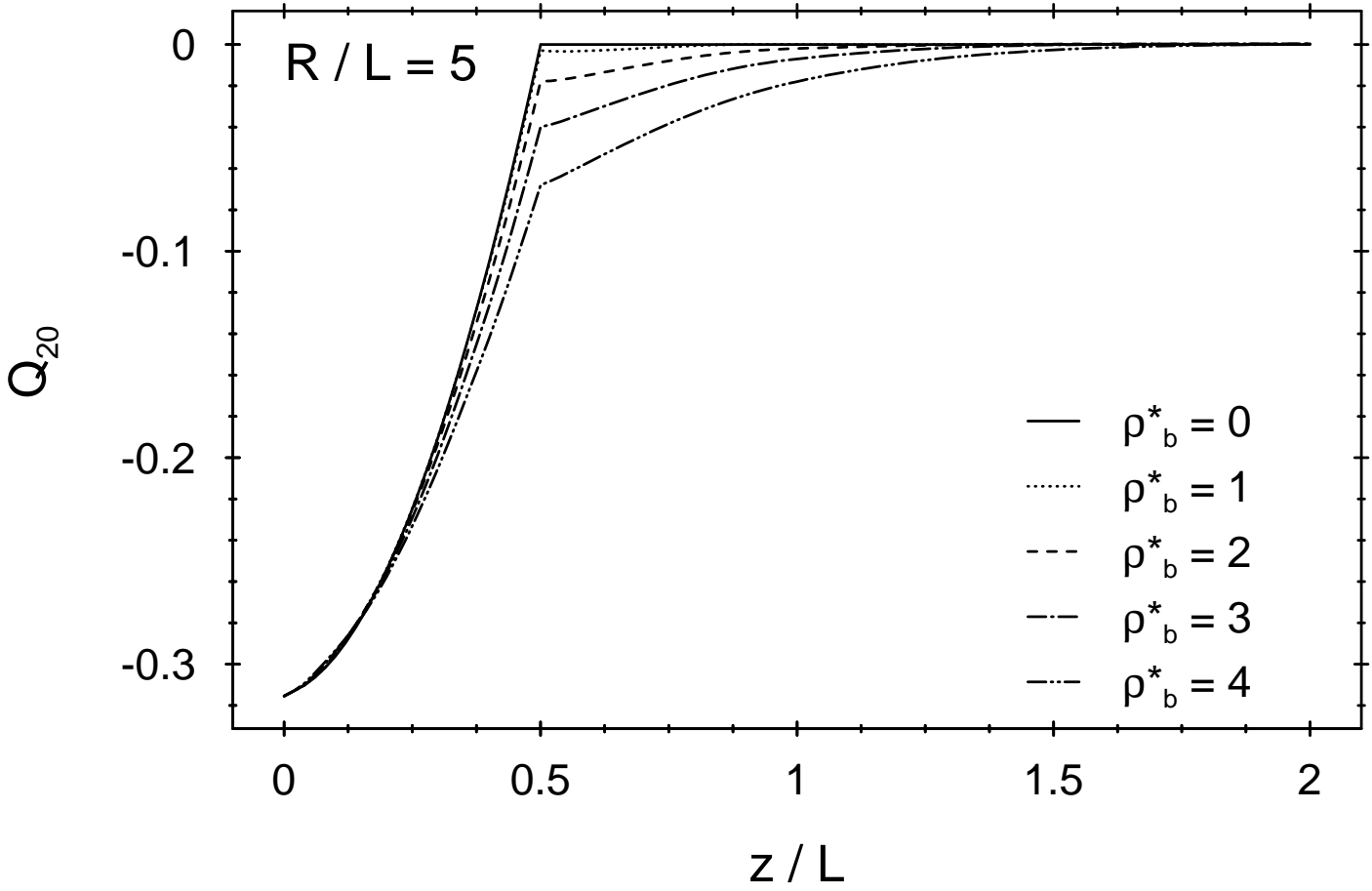
\begin{figure}
\caption{Order parameter $Q_{20}$ at fixed density $\rhos_b=2$ for
different wall geometries. Parallel orientations, i.e., negative
values of $Q_{20}$ are more favored by
negative curvatures.}
\label{fig:poniqrad}
\end{figure}

We define position dependent orientational order parameters as
\begin{equation} \label{Qlm}
  Q_{lm}(z)=\frac{1}{\rho_b n(z)} \int d\omega Y_{lm}^\ast(\omega)
  \hat\rho(z,\omega).
\end{equation}
For the sphere due to the azimuthal
symmetry one has $Q_{lm}=0$ for $m\neq 0$ . The lowest non-trivial order parameter $Q_{20}(z)$ is
plotted in Figs.~\ref{fig:poniqrho} and \ref{fig:poniqrad}. As
$\hat\rho(z,\omega)/(\rho_b n(z))\simeq 1$ for $z\to 0$ and in this
limit only
$\theta=\pi/2$ is allowed it follows that $Q_{20}(z\to
0)=-\sqrt{5/8\pi}=-0.3154$. Negative values of $Q_{20}$ indicate that
the rods are preferentially aligned parallel to the surface which is of
course enforced by the wall. The inter-particle interactions tend to align the
rods also for $z>L/2$, where they cannot directly touch the wall, and
increase the alignment for $z<L/2$ (see \abbref{fig:poniqrho}). The
alignment is stronger for positive than for negative curvature (see
\abbref{fig:poniqrad}). 

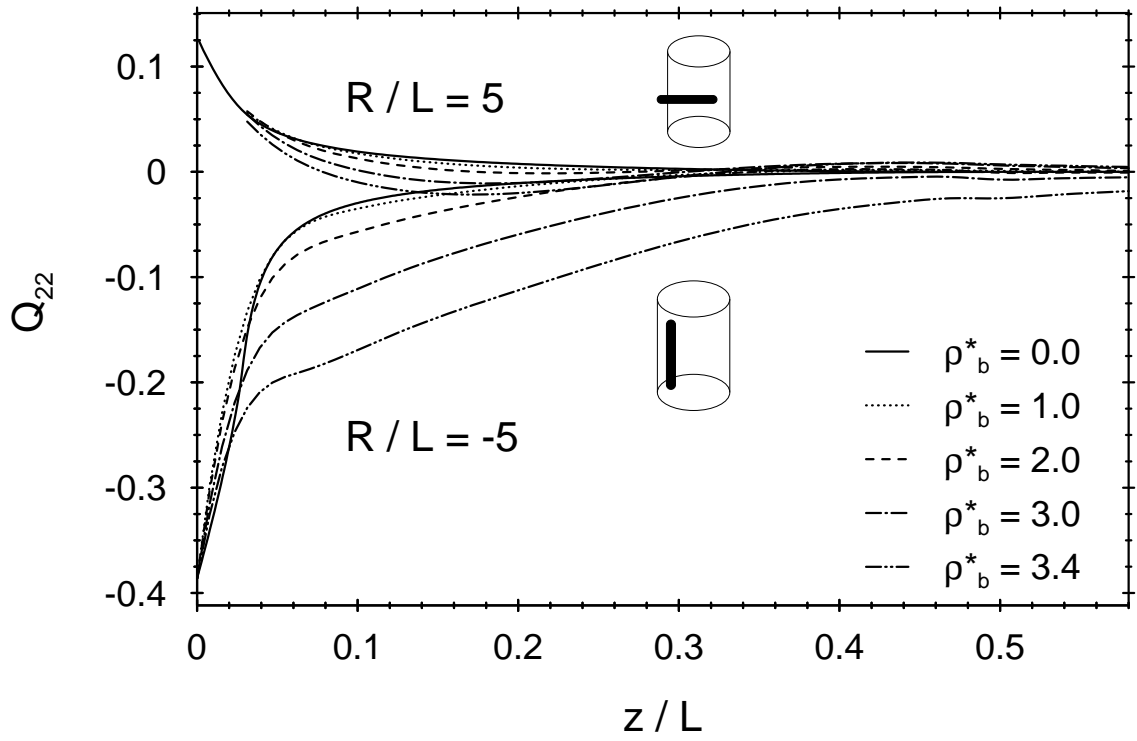
\begin{figure}
\caption{Biaxial order parameter $Q_{22}$ (see \eqpref{Qlm}) for
cylindrical walls of positive and negative curvature at different bulk
densities. The most probable orientations for positive and negative
values of $Q_{22}$ are indicated in the sketches. The strong increase
of the decay length with density for $R/L=-5$ signals  the formation of a nematic wetting layer.}
\label{fig:poniq22}
\end{figure}

The biaxiality of the orientational distribution at a cylinder is
measured by $Q_{22}(z)$. Positive (negative) values correspond to a preferential
orientation perpendicular (parallel) to the cylinder axis. If the orientational distribution is sharply peaked at
$\theta=\pi/2$ and $\phi=0$ (or $\phi=\pi/2$) $Q_{22}$ takes on its
maximum (minimum) value $Q_{22}=\pm\sqrt{15/32\pi}=\pm 0.3863$. The results
for cylinders of radius $R/L=\pm5$ are displayed in
\abbref{fig:poniq22}. As expected, particles {\it inside} a cylinder orient
themselves mainly parallel to the cylinder axis, the more the stronger
the interactions are. With increasing bulk density the decay of
$Q_{22}$ towards the bulk value 0 becomes
significantly slower, which probably signals the onset of
the formation of a nematic wetting layer upon approaching the
isotropic-nematic transition. This interpetration is supported by the fact that the iterations did not converge for
$\rhos_b\gtrsim3.5$. For this density range spontaneous biaxial orientational order has also been
predicted at a {\it planar} surface 
\cite{Poni:93}. For positive curvature  $Q_{22}$ has the opposite
sign and a much lower absolute value. The limit $z\to 0$ for $Q_{22}$ cannot be
determined rigorously in this case because the allowed region in $\omega$ space
does not reduce to a single point so that $Q_{22}(z\to 0)$ still depends on
an unknown function of $\phi$. Only in the ideal case $\rho_b\to 0$ this function is
constant so that $Q_{22}(z\to 0)=\sqrt{5/96\pi}=0.1288$. Here, the effect of
the interactions is to increase the probability for orientations
parallel to the axis so that $Q_{22}$ may even become negative.

%%%%%%%%%%%%%%%%%%%%%%%%%%%%%%%%%%%%%%%%%%%%%%%%%%%%%%%%%%%%%%%%%%%%%%
\section{Surface tension} \label{surf}

The curvature dependent surface tension $\gamma(R)$ is that 
contribution to the grand canonical potential which scales with the
surface area of the confining wall. In order to determine this quantity for
positive curvature one has to consider systems of finite size
$\cal L$ in the radial direction. However, such
systems necessarily contain a second, isotropic liquid - vacuum
interface generated by
the cutoff yielding the corresponding artificial surface contribution
$\gamma_{vac}$. Hence for a sphere we have
\begin{equation}  \label{gamdef}
  4\pi R^2 \gamma^S(R)=\lim_{{\cal L}\to\infty} \left[ \Omega(R,{\cal
  L})-\frac{4\pi}{3}((R+{\cal L})^3-R^3) \omega_b -4\pi(R+{\cal L})^2
  \gamma^S_{vac}(R+{\cal L}) \right]
\end{equation}
where $\omega_b=-p$ (see \eqref{omb}) is the bulk grand canonical
potential density and $p$ the bulk pressure.

The vacuum surface tension can be obtained separately by considering the
one-surface problem of a sphere of radius $R+{\cal L}$ filled
completely with an
isotropic fluid and in contact with the vacuum. In this case the
grand canonical potential is the sum of a bulk term and a surface term
proportional to $\gamma_{vac}$ which leads to
\begin{equation}
  4\pi(R+{\cal L})^2\beta\gamma^S_{vac}(R+{\cal L})=-\frac{1}{8}
  \rho_b^2 \int\limits_0^{R+{\cal L}} dr r \int\limits_{R+{\cal L}}^\infty dr' r'
  w_{00}(r,r').
\end{equation}
In the deriving this expression we have used the relation
\begin{equation} \label{w00v0}
  \frac{1}{4\pi r} \int\limits_0^\infty dr' r' w_{00}(r,r')=4 v_0,
\end{equation}
which can be proven using Eqs.~\klref{wS} and \klref{v0}. After some
algebra one finally obtains
\begin{eqnarray} \label{gamsph}
  4\pi R^2 \beta\gamma^S(R) & = & 4\pi \int\limits_{R}^\infty dr r^2 \left\{
  2\pi \int\limits_{-1}^1 dx \hat\rho(r,x)\left[ \ln 4\pi
  \hat\rho(r,x)\lambda^3 -1-\beta\mu \right]-\rho_b (\ln\rho_b
  \lambda^3 -1-\beta\mu) \right\} \nn\\
  & &+\half \rho_b^2 \int\limits_R^\infty dr_1 r_1 \left[ -4\pi r_1 v_0+\int\limits_R^\infty dr_2 r_2
  \sum_{l_1,l_2} \alpha_{l_1}(r_1) \alpha_{l_2}(r_2)
  w_{l_1l_2}(r_1,r_2) \right].
\end{eqnarray}

Inside a sphere the fluid volume is finite which does not allow to
carry out the thermodynamic limit. Instead we define $\gamma^S$ as
\begin{equation}
  4\pi R^2 \gamma^S(R)=\Omega(R)-\frac{4\pi}{3}|R|^3 \omega_b.
\end{equation}
The resulting expression for $\gamma(R)$ is identical to
\eqref{gamsph}, but with all radial integrations spanning the interval
from 0 to $|R|$
instead of from $R$ to $\infty$. If one uses the fact that the equilibrium
profile solves the Euler-Lagrange equation \eqref{ELGbare} and that the
bulk density satisfies the equation $\ln\rho_b\lambda^3=\beta\mu-\rho_b v_0$ these
results can be cast into the following simpler form:
\begin{eqnarray} \label{gamsphmin}
 4\pi R^2 \beta\gamma^S(R) & = & 4\pi \int\limits dr r^2[1-2\alpha_0(r)]
 \\
 & &{}-
 \half \rho_b^2 \int\limits dr_1 r_1 \left[ -4\pi r_1 v_0+\int\limits dr_2 r_2
  \sum_{l_1,l_2} \alpha_{l_1}(r_1) \alpha_{l_2}(r_2)
  w_{l_1l_2}(r_1,r_2) \right] \nn
\end{eqnarray}
with different integration limits for the outside and inside, as
stated above.
However, one should keep in mind that \eqref{gamsph} represents the surface
contribution to the density functional and is minimized by the
equilibrium profile whereas \eqref{gamsphmin} only applies to the
equilibrium solution. In practice the calculation of the surface
tension via both formulas provides a helpful check of the numerical
procedure.

The corresponding expression for the cylinder is
\begin{equation} 
 2\pi R \gamma^C(R)=\lim_{{\cal L}\to\infty} \left[ \Omega(R,{\cal L})/H
 -\pi ((R+{\cal L})^2-R^2)\omega_b -2\pi(R+{\cal L})
 \gamma^C_{vac}(R+{\cal L}) \right].
\end{equation}
No truncation in the axial direction is needed because $\Omega$ is
evidently proportional to the macroscopic height $H$. Here we could
confirm the analogue of \eqref{w00v0}, i.e.,
\begin{equation} 
  \int\limits_0^\infty dr' r' w_{0000}(r,r')=8\pi^2 v_0,
\end{equation}
only  numerically. The
resulting expressions for the surface tension, which with the 
modifications quoted above are also valid for the inside of a cylinder, are
\begin{eqnarray} \label{gamcyl}
  2\pi R \beta\gamma^C&=& 2\pi \int\limits_R^\infty dr\,r \left\{ \int\limits d\omega
  \hat\rho(r,\omega) \left[\ln 4\pi\hat\rho(r,\omega)-1-\beta\mu
  \right]
  -\rho_b (\ln\rho_b
  \lambda^3 -1-\beta\mu) \right\} \\
  & &+\half \rho_b^2 \int\limits_R^\infty dr_1 r_1 \left[-2\pi
  v_0+\int\limits_R^\infty  dr_2 r_2
  \sum_{l_1,l_2,m_1,m_2} \mu_{l_1m_1}(r_1) \mu_{l_2m_2}(r_2)
  w_{l_1m_1l_2m_2}(r_1,r_2) \right] \nn
\end{eqnarray}
and, at equilibrium,
\begin{eqnarray} \label{gamcylmin}
  2\pi R \beta\gamma^C&=& 2\pi \int\limits_R^\infty dr\, r
  \rho_b\left[1-\sqrt{4\pi} \mu_{00}(r)\right] \\
  & &-\half \rho_b^2 \int\limits_R^\infty dr_1 r_1 \left[-2\pi
  v_0+\int\limits_R^\infty  dr_2 r_2
  \sum_{l_1,l_2,m_1,m_2} \mu_{l_1m_1}(r_1) \mu_{l_2m_2}(r_2)
  w_{l_1m_1l_2m_2}(r_1,r_2) \right]. \nn
\end{eqnarray}

We remark that the surface tension depends on the assumed position of
the actual surface \cite{Lajtar:87}, i.e., on the definition of what
is denoted as the volume of the
sphere or cylinder, which is not uniquely determined. A different
choice for this position would alter the volumes and surface areas
occuring in \eqref{gamdef} and thereby in general lead to a different
value of $\gamma$. For the planar case this change is
$\Delta\gamma=p\Delta x$, where $\Delta x$ is the shift in the surface
position, while more complicated relations emerge for curved surfaces,
which may significantly change the curvature
dependence of what is denoted as the surface tension. On the other hand, experimentally observable quantities
do not depend on this arbitrariness of assigning a particular value to
the volume of the cavity. For the thin rods we employed
the natural definition that the defining surface is given by the
position of the rod ends at closest approach. But already for hard
spheres, or for rods of finite thickness, there are at least two
possible ``natural'' definitions (see, e.g., Figs.~1 and 14 in
\cref{Goetzelmann:96}). 

The surface tension $\gamma_0$ in the ideal gas limit 
is obtained from Eqs.~\klref{gamsphmin} and \klref{gamcylmin} by
neglecting the
interaction contributions  and by inserting the ideal profiles
$\hat\rho_0(r,\omega)$ from Appendix~\ref{rhoid}. This results in
\begin{equation} \label{gam0sph}
  \frac{\beta\gamma^S_0}{\rho_0}=\frac{1}{R^2} \int\limits dr\, r^2 (1-2\alpha_0(r))
  =\frac{1}{R^2} \int\limits dr\, r^2 (1-x_{max}(r))
\end{equation}
for a sphere (with integration limits for $R\lessgtr 0$ as described above) and
\begin{equation} \label{gam0cyl}
  \frac{\beta\gamma^C_0}{\rho_0}=\left\{
  \begin{array}{ll}
  \frac{1}{R} \int\limits_R^\infty dr\, r \left[1-\frac{2}{\pi} \int\limits_0^1 dx
  \phi_c(r,x) \right], & R>0 \\
  \frac{1}{|R|} \int\limits_0^{|R|} dr\, r \left[1- \int\limits_0^1 dx
  (1-\frac{2}{\pi} \phi_c(r,x)) \right], & R<0 
  \end{array}
  \right.
\end{equation}
for a cylinder. This limit has been discussed extensively by Yaman
et al. \cite{Yaman:97:PRL,Yaman:97}, who found the surprising result
$\beta\gamma_0/\rho_0=L/4$ for rods outside {\it any} convex body,
while this expression is modified inside a sphere or a cylinder, so that
\begin{equation} \label{gamSid}
  \frac{\beta\gamma^S_0}{\rho_0}=\left\{\begin{array}{ll}
   \frac{L}{4}-\frac{L^3}{48 R^2}, & R<0 \\
   \frac{L}{4}, & R>0
				 \end{array}\right.
\end{equation}
and 
\begin{equation} \label{gamCid}
 \frac{\beta\gamma^C_0}{\rho_0}=\left\{
  \begin{array}{ll}
    \frac{L}{4}-\frac{L^3}{128 R^2}+O(R^{-3}), & R<0 \\
   \frac{L}{4}, & R>0
  \end{array}\right..
\end{equation}
We have confirmed these results (analytically for the sphere,
numerically for the cylinder) by using Eqs.~\klref{gam0sph} and
\klref{gam0cyl}. The most interesting aspect of these findings is that
the surface tension is not analytical at $1/R=0$ which may lead to
unexpected behavior of membranes immersed in colloidal rod
suspensions.

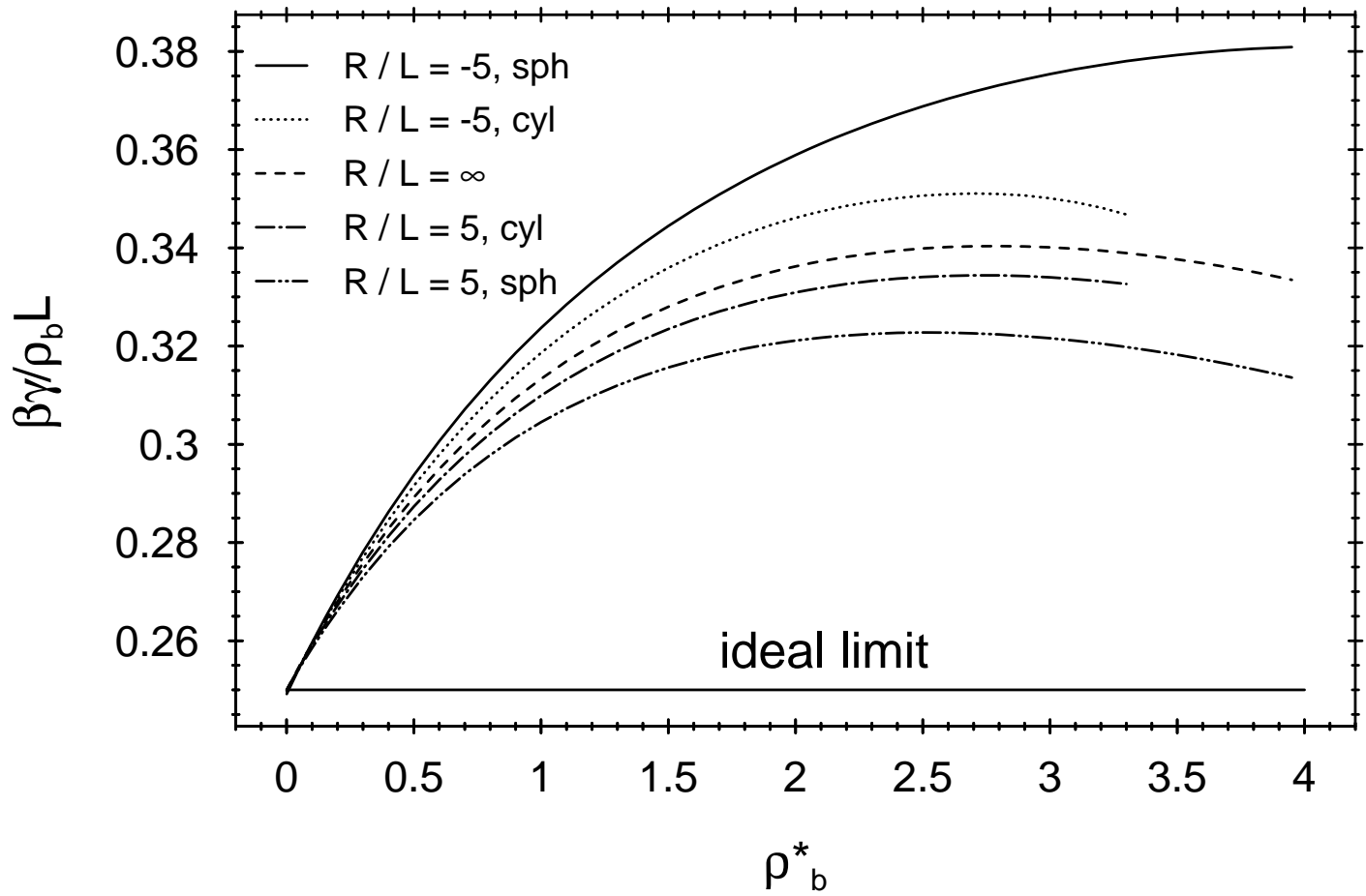
\begin{figure}
\caption{Density dependence of the surface tension for different wall
geometries and curvatures. In the ideal limit $\beta\gamma/\rho_b L$
takes on the density independent values 0.25 for $1/R\geq 0$
(cylinders, spheres, and planar wall), and 0.2492 (0.2497) for a sphere (cylinder) with
$R/L=-5$. The differences between these three values are not visible
on the scale of the figure.}
\label{fig:surfradf}
\end{figure}

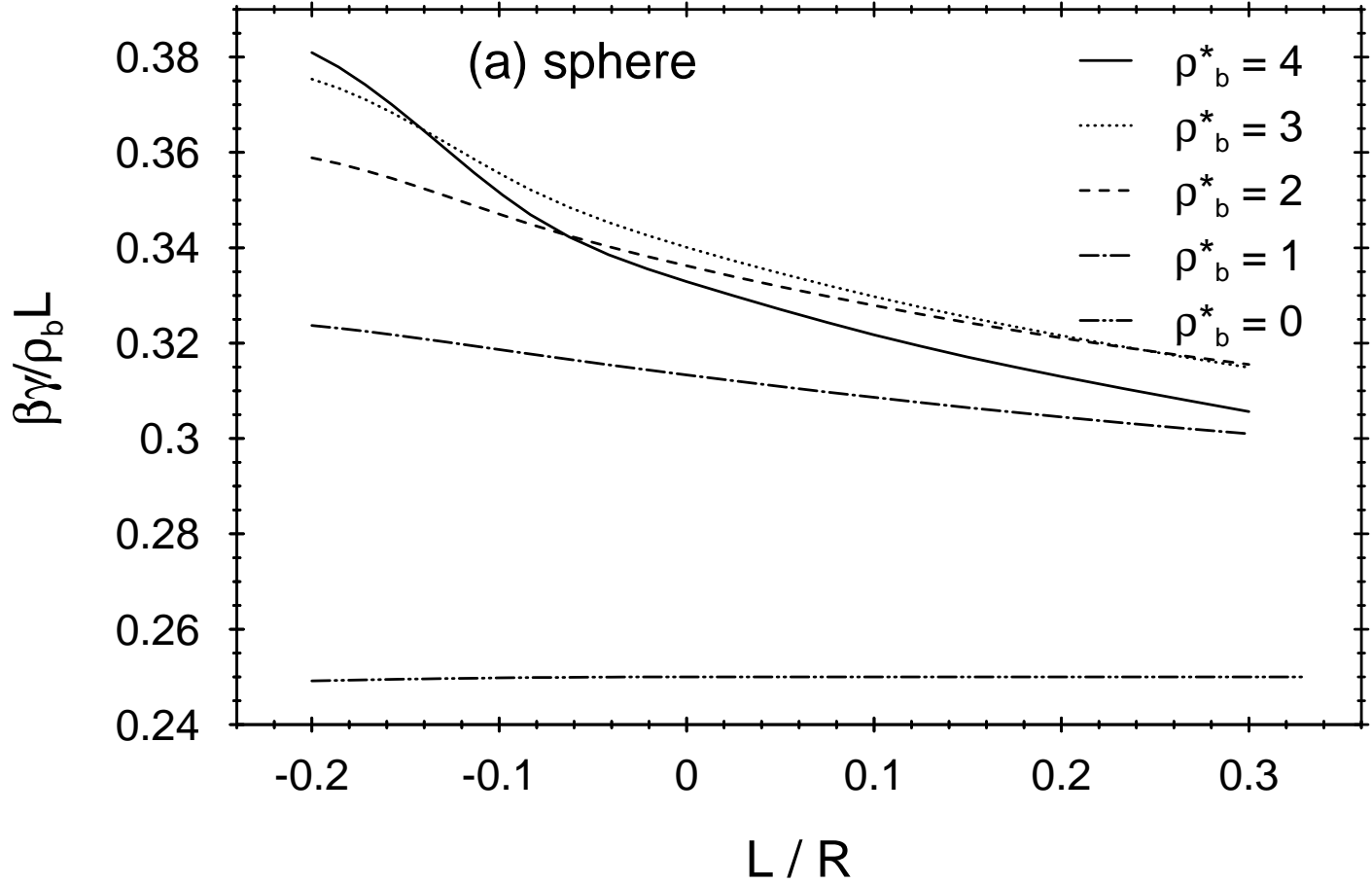
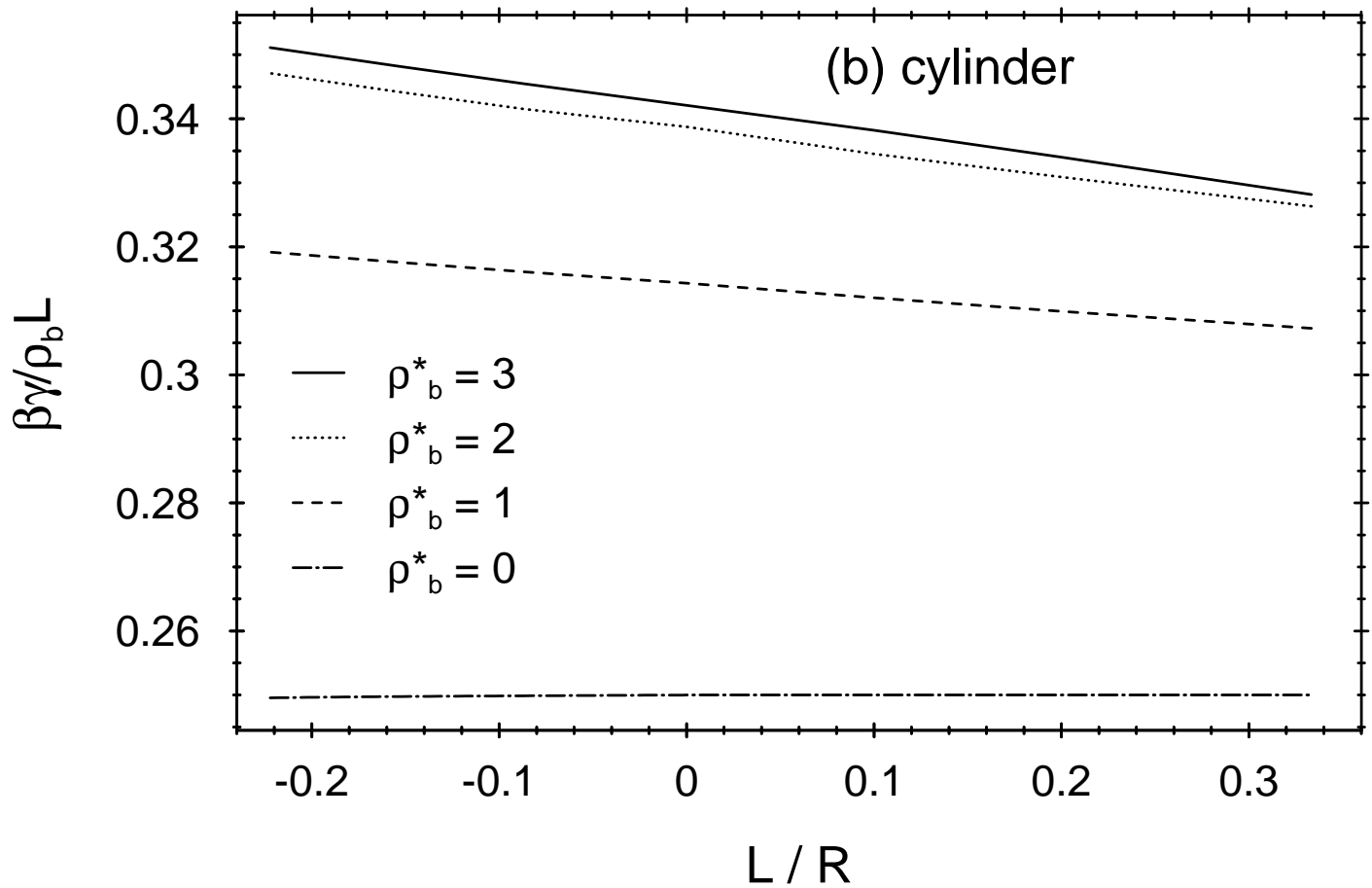
\begin{figure}
\caption{Curvature dependence of the surface tension for (a) spheres and
(b) cylinders at various densities. In the limit of non-interacting rods
($\rho_b=0$) there is a slight decrease of $\gamma$ for negative
curvature, which is hardly visible on this scale. Taking into account
the interparticle interactions leads to a substantial enhancement of the
curvature dependence already for $\rhos_b=1$. We have interpolated
smoothly between the 7 data points calculated for each curve.}
\label{fig:surfrhof}
\end{figure}

We have determined the surface tension in the presence of the
interparticle interactions for a series of
bulk densities and radii. The results are shown in
\abbref{fig:surfradf} as function of $\rho_b$ for fixed $R$ and in
\abbref{fig:surfrhof} as function of $R$ for fixed $\rho_b$. For each
data point four (spheres) or three (cylinders) numerical calculations with different values
of the cutoff $l_{max}$ were performed. The results were extrapolated
to $l_{max}=\infty$ using a fit function linear or quadratic in
$1/l_{max}$. The differences betweeen the values at the largest
$l_{max}$ and the extrapolation become considerable [$\Delta
(\beta\gamma)/(\rho_b L)\simeq 0.02$] at large densities. From comparison
of the results obtained by quadratic and linear extrapolation we
estimate the error of $\beta\gamma/(\rho_b L)$ to be 0.01 for $\rhos_b=4$
but only 0.001 for $\rhos_b\leq2$. Finally we have interpolated
smoothly between the 13 data points taken for each radius. In the case
of a cylinder at the highest density $\rhos_b=3$ the result obtained
separately for the special case of a
planar wall lies slightly (by 0.003 in the units used here) above the
almost linear curve through the other points. A possible explanation
for this observation
is that at the planar wall a uniaxial orientational distribution has been
assumed while the actual equilibrium profile might exhibit a small
spontaneous biaxiality as found at the cylindrical walls. Therefore
the planar wall result has not been used for the interpolation scheme in this
case.

Figure~\ref{fig:surfradf} shows the surface tension divided by the
density to allow for a better comparison with the ideal rods results
that predict a density independent constant for this ratio. In all
cases the interaction significantly increases this quantity, by up to
50\% for the intermediate density $\rhos_b=2$. In the upper density
range saturation or the formation of a maximum are observed. The results for a planar
wall are in good agreement with those obtained by Mao et al. \cite{Mao:97}, who
effectively used the same theory but a different numerical method. For
almost all densities the surface tension is higher for negative than
for positive curvature, in contrast to the behavior at $\rho_b\to 0$. In
the latter case $\gamma$ is constant for $R>0$; the slight $R$
dependence for $R<0$ is hardly visible on the scale of
\abbref{fig:surfrhof}. On the other hand, for  densities of the
order of 1 in reduced units the dependence on $R$ is dominated by a term
linear in $1/R$ that is absent in the ideal limit. The dependence on $R$
becomes stronger and less linear with increasing density, especially
for the spherical case.

Due to the limited number of radii for which calculations were
performed we are not able to decide whether the small discontinuity
of the second derivative of $\gamma(1/R)$ at $1/R=0$ that occurs in
the ideal limit persists also at finite bulk densities. However, our
data do not preclude this possibility. It is commonly assumed that the
surface free energy density for a general surface with principal
curvatures $1/R_1$ and $1/R_2$ has the Helfrich form
\cite{Helfrich:73}
\begin{equation}
  \gamma(1/R_1,1/R_2)=\gamma^P+2\kappa
  \left(\half(\frac{1}{R_1}+\frac{1}{R_2})
  -c_0\right)^2+\bar\kappa\frac{1}{R_1R_2}+\cdots
\end{equation}
with the stiffness coefficients $\kappa$ and $\bar\kappa$ and the
spontaneous curvature $c_0$. Inter alia, this form predicts that the contribution
to $\gamma(1/R)$ linear in $1/R$ for a sphere is twice that for a
cylinder. From our numerical results we found that this relation is
approximately fulfilled at low densities, but there are substantial
deviations from it at higher densities. Moreover, the Helfrich expression
cannot be applicable for all signs of the curvatures already in the
ideal limit due to the aforementioned non-analyticity of $\gamma^S(1/R)$
and $\gamma^C(1/R)$.

We conclude that for phenomena for which the curvature dependence of the
surface tension matters the non-interacting approximation is
qualitatively wrong already at relatively low densities (compared to
the transition density to the nematic phase) and at any
fixed curvature it produces rather
large quantitative errors in the absolute value of $\gamma$.

%%%%%%%%%%%%%%%%%%%%%%%%%%%%%%%%%%%%%%%%%%%%%%%%%%%%%%%%%%%%%%%%%%%%%%
\section{Summary} \label{summary}

For an isotropic fluid of needle-like hard spherocylinders of length
$L$ near hard spherical or cylindrical walls we have obtained the
following main results:

\begin{enumerate}
\item Due to the interparticle interactions the probability of finding
a particle with a given orientation is strongly increased close to the
wall when compared to the bulk fluid. At a given distance $z$ of the
center of mass from the
wall orientations for which one end of the rod touches the wall are
most favorable (see \abbref{fig:rhotld}).

\item Since the range of accessible orientations decreases when the
particle approaches the wall the orientationally averaged density
vanishes for $z\to 0$. It exhibits a cusp at $z=L/2$ where the
rods lose contact with the surface (\abbref{fig:ponin}).

\item The parallel alignment favored by the surface decays more slowly
when the bulk density is increased (\abbref{fig:poniqrho}) and is
stronger if the wall curves towards the fluid instead of away from
it (\abbref{fig:poniqrad}).

\item A cylindrical wall curving towards the fluid induces biaxial
orientational order with preferential alignment parallel to the
cylinder axis. With increasing bulk density a nematic wetting layer
develops in this case (see \abbref{fig:poniq22}).

\item The density and curvature dependences of the wall-fluid surface
tension are shown in Figs.~\ref{fig:surfradf} and
\ref{fig:surfrhof}. In contrast to the results for non-interacting
particles given by Eqs.~\klref{gamSid} and \klref{gamCid} the surface
tension decreases with increasing curvature and exhibits a linear
behavior around the planar limit which leads to a spontaneous
curvature of a membrane away from a fluid of rod-like (colloidal)
particles.

\end{enumerate}

%%%%%%%%%%%%%%%%%%%%%%%%%%%%%%%%%%%%%%%%%%%%%%%%%%%%%%%%%%%%%%%%%%%%%%
%%%%%%%%%%%%%%%%%%%%%%%%%%%%%%%%%%%%%%%%%%%%%%%%%%%%%%%%%%%%%%%%%%%%%%
\begin{appendix}

\section{Determination of the expansion coefficients of the Mayer
function} \label{fllm}

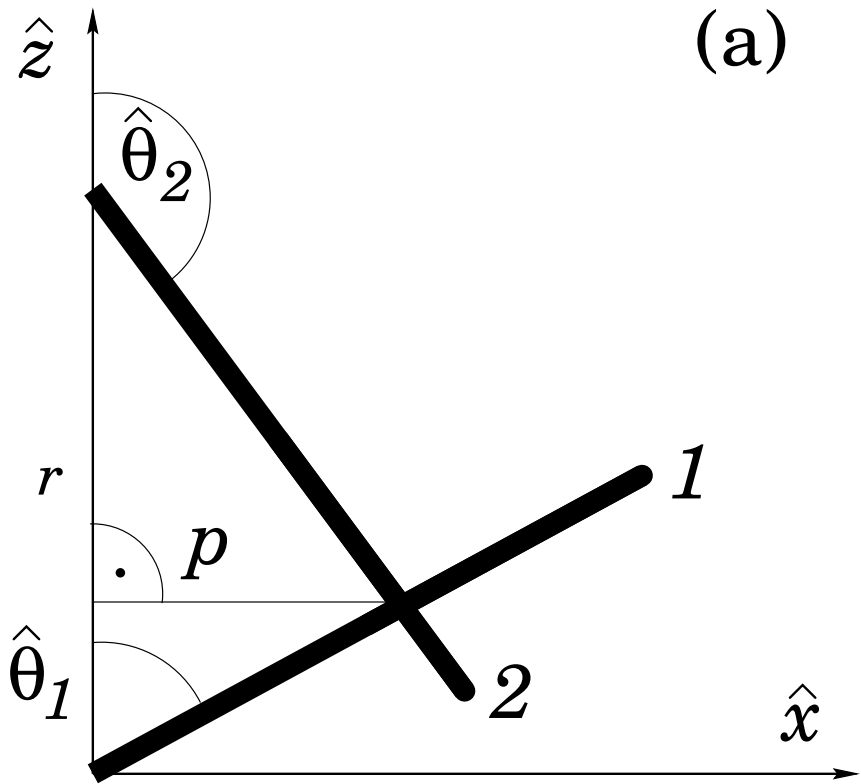
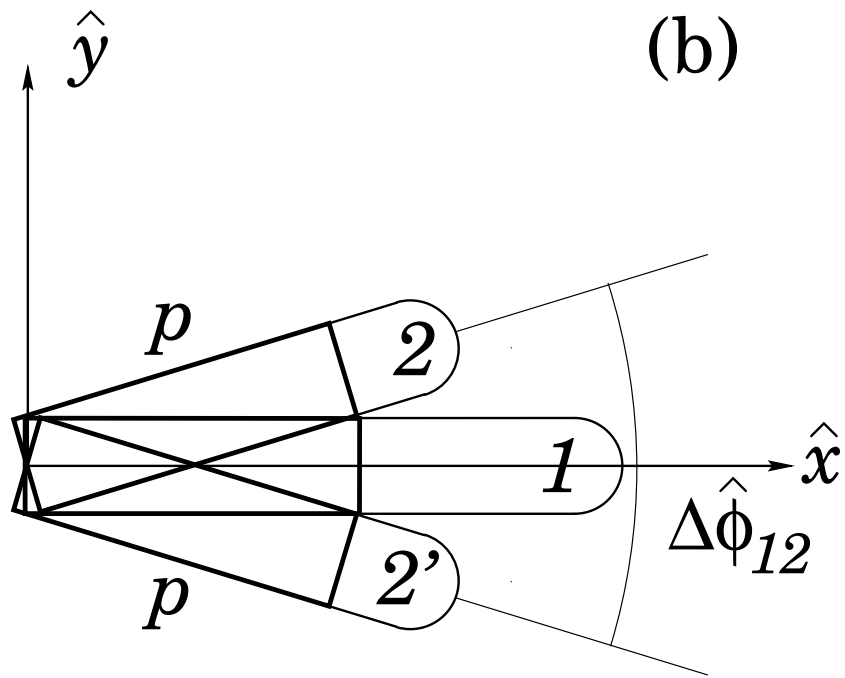
\begin{figure}
\caption{Geometry of two overlapping rods in the limit $D/L\to
0$. Only one half of each rod is shown. Their centers lie on the $\hat
z$ axis and have a distance $r$. In
order to overlap both rods must lie approximately in the same 
plane (a). Part (b) shows a vertical projection from which the allowed
azimuthal range $\Delta\hat\phi_{12}=2 D/p$ can be determined. Rod 2
is drawn in the two positions for which it just touches rod 1.}
\label{fig:overlap}
\end{figure}

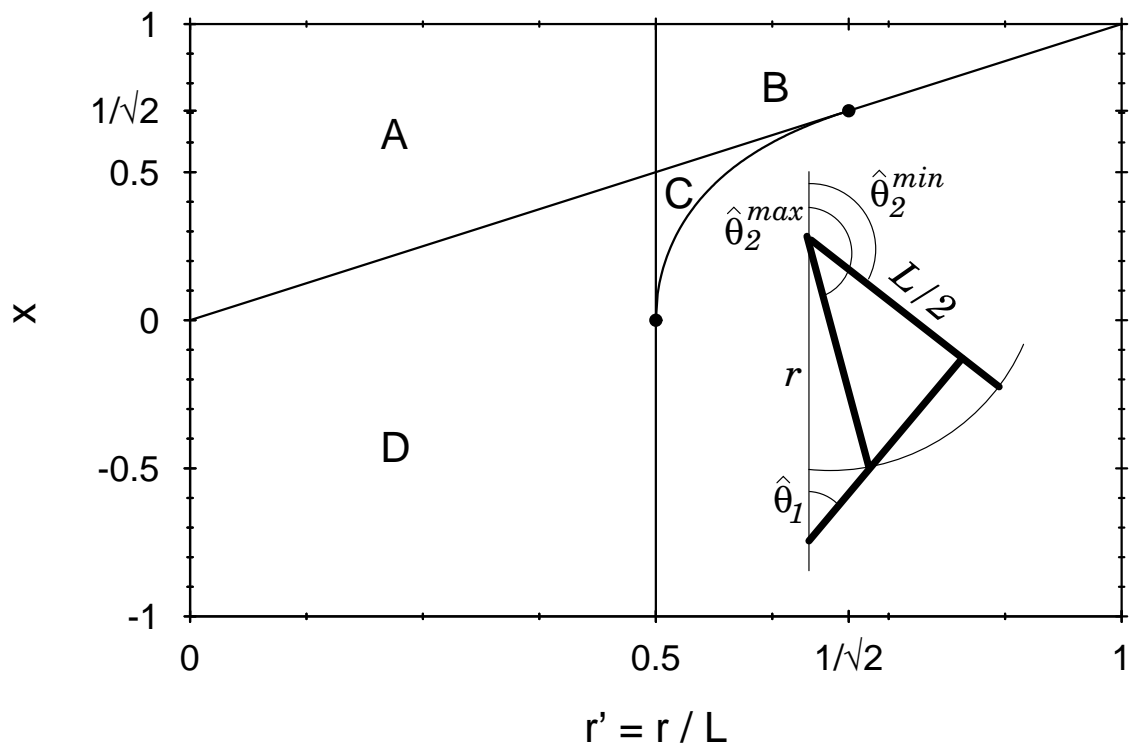
\begin{figure}
\caption{Illustration of the cases that must be distinguished for the
determination of the minimum ($x_{min}=\cos\hat\theta_2^{max}$) and
maximum ($x_{max}=\cos\hat\theta_2^{min}$) value of $x_2$ so
that two rods overlap for given $x_1=\cos\theta_1$ and $r$. The appropriate
expressions for $x_{min}$ and $x_{max}$ for the regions A, B, C, and D are given in
the main text. In the unlabeled region containing the inset overlap is not possible. The inset
shows an example from region B where the end of rod 2 touches rod 1 at
the minimum angle and vice versa at the maximum angle.}
\label{fig:xminxmax}
\end{figure}

Equation \klref{fllmdef} serves as the starting point which defines
the expansion coefficients $f_{l_1l_2m}(r)$. As far as the azimuthal
angles are concerned the integrand depends only
on $\hat\phi_{12}=\hat\phi_2-\hat\phi_1$ because
\begin{equation}
  Y_{lm}(\theta,\phi)=k_{lm} P_{lm}(\cos\theta) e^{-im\phi},
\end{equation}
where the coefficients $k_{lm}$ relating the spherical harmonics to the
associated Legendre functions $P_{lm}$ are given by
\begin{equation}
  k_{lm}=(-1)^m \left(
  \frac{2l+1}{4\pi}\frac{(l-m)!}{(l+m)!}\right)^{1/2}
  \mbox{ for } m\geq 0, \qquad
  k_{l\mq}=(-1)^m k_{lm}.
\end{equation}
Therefore after the substitutions
$\hat\phi_s=\half(\hat\phi_1+\hat\phi_2)$ and
$\hat\phi_{12}=\hat\phi_2-\hat\phi_1$ the integral over $\hat\phi_s$
renders a factor $2\pi$.
In \eqref{fllmdef} we now perform the integrations in the following order:
\begin{eqnarray}
  f_{l_1l_2m}(r)&=&2\pi k_{l_1m} k_{l_2\mq} \int_{-1}^1 d\cos\hat\theta_1 \int_{-1}^1
  d\cos\hat\theta_2 \int_0^{2\pi} d\hat\phi_{12}\,
  f(r,\hat\theta_1,\hat\theta_2,\hat\phi_{12}) P_{l_1m}(\cos\hat\theta_1) \\
  & &{}\times
  P_{l_2m}(\cos\hat\theta_2) e^{i m \hat\phi_{12}} \nn.
\end{eqnarray}
In the limit $D/L\to 0$ the two rods overlap only in a small range
$\Delta\hat\phi_{12}$ around $\hat\phi_{12}=0$ or $\hat\phi_{12}=\pi$,
approximately given by $\Delta\phi_{12}=2 D/p$ where $p$ is the
distance between the intersection point and the line joining the
centers of the rods, as shown in \abbref{fig:overlap}. From this
figure one easily derives $p=r/(\cot\theta_1-\cot\theta_2)$; based on
appropiately modified figures one finds that this expression is also valid
if $\theta_1>\pi/2$ or $\theta_2<\pi/2$. Therefore it is sufficient to
replace the $\hat\phi_{12}$ integration by the factor $\Delta\hat\phi_{12}$,
to replace the integrand by its value at $\hat\phi_{12}=0$ or $\pi$,
and to examine only the overlap of two infinitely thin rods in the
$\hat x\hat z$ plane. In integrations over the full spatial angles
$\hat\omega_i$ we assign an intrinsic
directionality to the rods such that their ``front ends'' point into the direction
$(\hat\theta_i,\hat\phi_i)$ and their ``rear end'' into the direction
$(\pi-\hat\theta_i,2\pi-\hat\phi_i)$. In order to calculate $f_{l_1l_2m}$ it is
sufficient to integrate over the configurations with overlap of the two front
ends, where necessarily $\hat\phi_{12}\simeq 0$. One can easily show
by appropriate substitutions that the other three possibilities for
overlap yield exactly the same contribution to the total integral, so
that one finds
\begin{eqnarray} \label{fllmcalc}
  f_{l_1l_2m}(r)&=&-16\pi k_{l_1m} k_{l_2\mq} \frac{D}{r} \int_{-1}^1
  dx_1\, P_{l_1m}(x_1) \\
 & & {}\times
  \int\limits_{x_{min}(r/L,x_1)}^{x_{max}(r/L,x_1)} dx_2 \,
  P_{l_2m}(x_2)
  \left(\frac{x_1}{\sqrt{1-x_1^2}}-\frac{x_2}{\sqrt{1-x_2^2}}
  \right). \nn
\end{eqnarray}
Here $x_{min}$ and $x_{max}$ denote the smallest and largest value of
$x_2=\cos\hat\theta_2$ for given $x_1=\cos\hat\theta_1$ and $r/L$ for which
the front halves of the rods overlap, as shown for an example in the
inset of \abbref{fig:xminxmax}. They can be determined by
tedious but straightforward geometry, which yields two basic formulas
for the $x$ value at {\it t}ouching,
depending on whether the end of rod 1 touches rod 2 or vice versa:
\begin{equation}
  x_{t}^{(1)}(r',x)=\frac{2r'-x}{\sqrt{1-2r'x+r'^2}} \qquad
  x_{t}^{(2,\pm)}(r',x)=2r'(1-x^2)\pm x\sqrt{1-4r'^2(1-x^2)}
\end{equation}
with $r'=r/L$. As illustrated in \abbref{fig:xminxmax} the following
regions in the $(r',x)$ plane must be distinguished:
\begin{itemize}
\renewcommand{\theenumi}{\Alph{enumi}}
 \item[A,] for $0\leq r'\leq 1/2$ and $r'\leq x\leq 1$: $x_{min}=x_{t}^{(1)}$ and
 $x_{max}=1$,
 \item[B,] for $1/2\leq r'\leq 1$ and $r'\leq x\leq 1$: $x_{min}=x_{t}^{(1)}$ and
 $x_{max}=x_{t}^{(2,+)}$,
 \item[C,] for $1/2\leq r'\leq 1/\sqrt{2}$ and $r\sqrt{1-1/4r'^2}\leq x\leq r'$: $x_{min}=x_{t}^{(2,-)}$ and
 $x_{max}=x_{t}^{(2,+)}$,
 \item[D,] for $0\leq r'\leq 1/2$ and $-1\leq x\leq r'$: $x_{min}=x_{t}^{(2,-)}$ and
 $x_{max}=1$.
\end{itemize}

For given $l$ and $m$ the integration over $x_2$ in \eqref{fllmcalc} can be
carried out analytically. In the remaining numerical integration special care
must be taken for $r'<1/2$ due to the square root singularity at
$x_1=1$. All coefficients $f_{l_1l_2m}$ are of the order of $D/L$ and vanish
for $r/L>1$. They diverge for $r/L\to 0$, but they appear only  in the
product $r f_{l_1l_2m}(r)$ (see Eqs.~\klref{wS}, \klref{wP}, and
\klref{wC}) which is finite in this limit. For two finite
values of $D/L$ the coefficients $f_{l_1l_2m}$ have been calculated by
Moore and McMullen \cite{Moore:90}.

A useful check of the numerical results is obtained from the observation that
the excluded volume $v_{ex}(\cos\gamma)$ for fixed angle $\gamma$
between the particle axes is related to the Mayer function via
$v_{ex}(\cos\gamma)=-\int d^3\rd\,f(\rv_{12},\omega_1,\omega_2)$ from
which one derives
\begin{equation}
  \int_{-1}^1 d\cos\gamma\, P_l(\cos\gamma)
  v_{ex}(\cos\gamma)=\frac{2}{2l+1} \sum_{m} (-1)^{m+1} \int_0^\infty
  dr\, r^2 f_{llm}(r).
\end{equation}
The left hand side can easily be determined using the well-known
result (see, e.g., \cref{Vroege:92}) $v_{ex}(\cos\gamma)=2 DL^2
|\sin\gamma|+O(D^2 L)$ and thus provides a sum rule for the second
moments of the expansion coefficients. Our numerical results passed
this check.

%%%%%%%%%%%%%%%%%%%%%%%%%%%%%%%%%%%%%%%%%%%%%%%%%%%%%%%%%%%%%%%%%%%%%%
\section{Density profiles for non-interacting rods} \label{rhoid}

In the ideal limit $\hat\rho(r,\omega)$ adopts the constant value
$\rho_0/4\pi$ for orientations $\omega$ that are allowed by the
presence of the hard wall and vanishes otherwise. Thus it is sufficient to
determine the limiting orientations for which the rod just touches the
wall. At a planar surface the maximum allowed value for $x=\cos\theta$
is $x_{max}=2 z/L$ for $z\leq L/2$ while the rod cannot touch the wall for
$z>L/2$ so that $x_{max}(z\geq L/2)=1$. At all surfaces the minimum value of $x$ clearly
is $-x_{max}$ because of the head-tail symmetry. Thus in the following
it is sufficient to consider positive $x$.

\subsection{Outside a sphere}

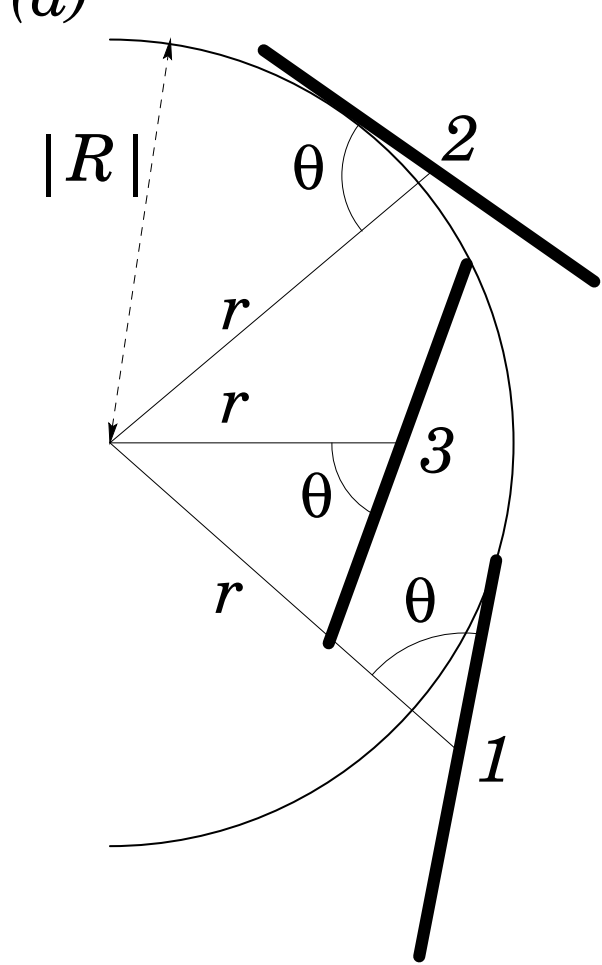
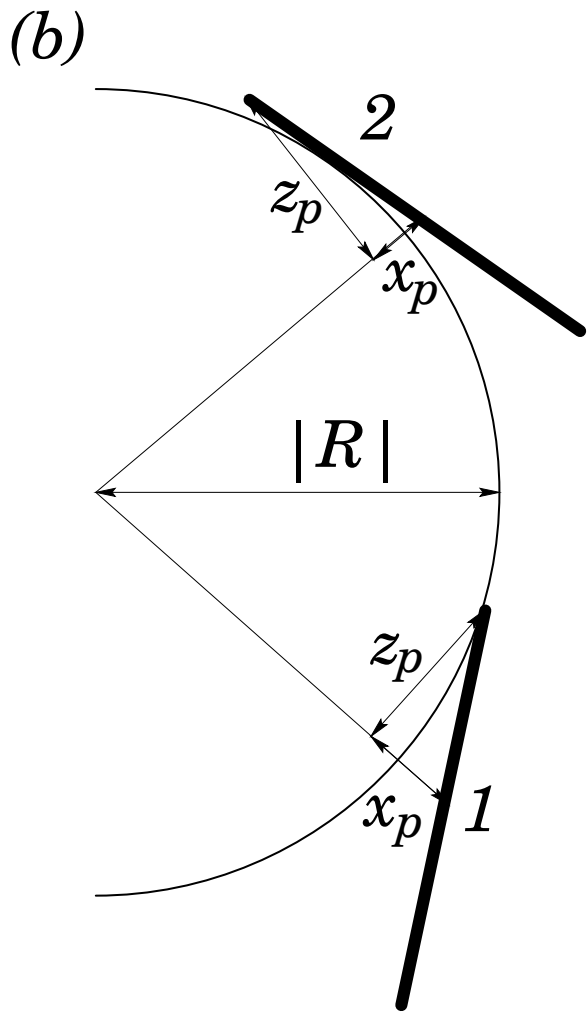
\begin{figure}
\caption{Geometries for the determination of the allowed orientations
at curved walls. (a) Outside a spherical wall the rod touches the wall
with its end if $r>r_c$ (rod 1), but tangentially if $r<r_c$ (rod 2).
Inside a sphere only end contact can occur (rod 3). From this
figure  Eqs.~(\pref{xmaxos}) and (\pref{xmaxis}) can be derived. (b)
Projection of the corresponding problem for a cylindrical wall onto
the plane perpendicular to the cylinder
axis. Equations~(\pref{xmaxoce}) and (\pref{xmaxoct}) can be derived
using $x_P=\frac{L}{2}\sin\theta \cos\phi_c$ and
$z_P=\frac{L}{2}\cos\theta$.}
\label{fig:wallcont}
\end{figure}

As illustrated in \abbref{fig:wallcont}(a),
there are two different ways how a rod can touch the outside of a
sphere:  when the
rod is sufficiently far from the surface its end touches the wall upon
rotation, whereas when it is close to the wall at contact it will
touch it tangentially. The crossover between these two regions
takes place at $r_c=\sqrt{R^2+L^2/4}$ and straightforward geometrical
reasoning yields
\begin{equation} \label{xmaxos}
x_{max}=\left\{
\begin{array}{cl}
 1, & r\geq R+L/2 \\
 {(L^2/4-R^2+r^2)}/{(r L)}, & r_c\leq r\leq R+L/2\\
 {\sqrt{r^2-R^2}}/{r}, & R\leq r\leq r_c \qquad.
\end{array}
\right.
\end{equation}

\subsection{Inside a sphere}

In this case there is a minimum distance $r_c=\sqrt{R^2-L^2/4}$ from
the surface beyond which all orientations are forbidden. In the
accessible region $|R|-L/2\leq r\leq r_c$ one finds [see \abbref{fig:wallcont}(a)]
\begin{equation} \label{xmaxis}
  x_{max}=\frac{R^2-L^2/4-r^2}{rL}.
\end{equation}

\subsection{Outside a cylinder}

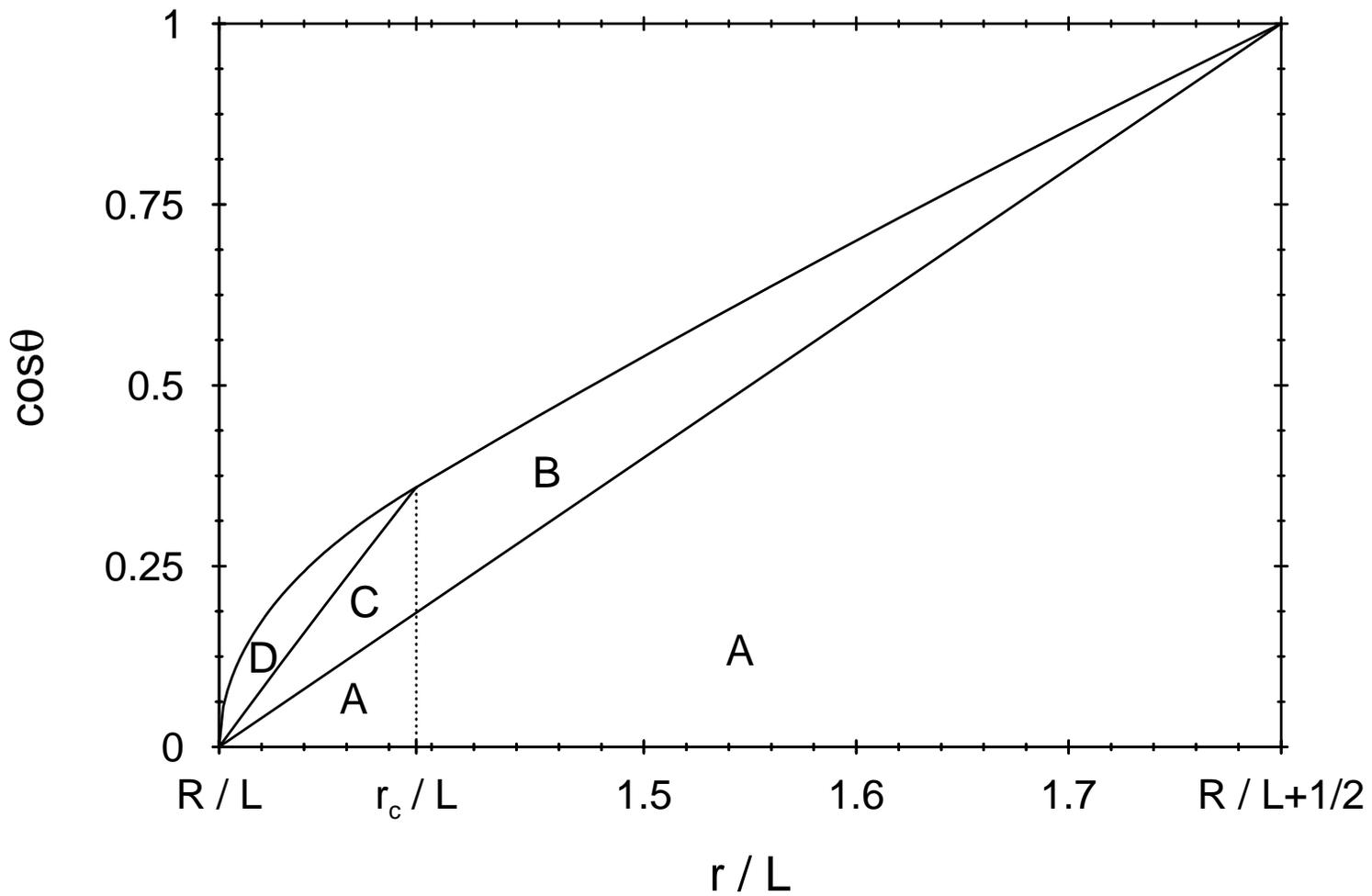
\begin{figure}
\caption{In calculating the accessible orientational space for a rod
outside a cylinder different expressions, presented in the main text,
apply in the regions labeled A, B, C, and D for the maximum azimuthal
angle $\phi_c$ at given values of $x=\cos\theta$ and $r$. The figure
corresponds to the case $R/L=1.3$, but its topology is the same for all
radii. Only for configurations corresponding to region D the rod touches the cylinder tangentially. For
 larger values of $R$ as actually used in the calculations presented above $r_c$ is
closer to $R$ and the regions C and D are much smaller.}
\label{fig:cylcase}
\end{figure}

For a cylinder the profile  depends in addition on the azimuthal angle $\phi$
which we always measure from the axis that is perpendicular to both
the cylinder axis and the surface normal. Thus here we have to determine the
range $I_\phi$ of allowed values of $\phi$ for fixed $r$ and $\cos\theta$. If the
rod touches the cylinder at the angle $\phi_{c}\in [0,\pi/2]$ then due
to  symmetry one has
$I_\phi=[0,\phi_c]\cup[\pi-\phi_c,\pi+\phi_c]\cup[2\pi-\phi_c,2\pi]$.
With the help of \abbref{fig:wallcont}(b) one finds for contact
between the rod end and the surface 
\begin{equation} \label{phim1} \label{xmaxoce}
  \cos\phi_c=\frac{2\sqrt{R^2-(r-L/2\cos\theta)^2}}{L \sin\theta}
\end{equation}
whereas for tangential contact one has
\begin{equation} \label{phim2} \label{xmaxoct}
  \cos\phi_c=\cot\theta \frac{R^2}{\sqrt{r^2-R^2}}.
\end{equation}
By determining the transitions between these two cases as well as
those to the ranges $I_\phi=\emptyset$ and $I_\phi=[0,2\pi]$ we are led to
distinguish the following cases (see \abbref{fig:cylcase}):
\begin{itemize}
\item[A,] $R\leq r\leq R+L/2$ and $0\leq \cos\theta\leq 2(r-R)/L$: $\cos\phi_c=0$,
\item[B,] $r_c=\sqrt{R^2+L^2/4}\leq r\leq R+L/2$ and
$2(r-R)/L\leq \cos\theta\leq (L^2/4-R^2+r^2)/(L r)$: $\cos\phi_c$ as given by
\eqref{phim1},
\item[C,] $R\leq r\leq r_c$ and $2(r-R)/L\leq \cos\theta\leq 2(r^2-R^2)/(L r)$:
$\cos\phi_c$ as given by \eqref{phim1},
\item[D,] $R\leq r\leq r_c$ and $2(r^2-R^2)/(L r)\leq \cos\theta\leq \sqrt{r^2-R^2}/r$:
$\cos\phi_c$ as given by \eqref{phim2}.
\end{itemize}
In the remaining region within $R\leq r\leq R+L/2$ all $\phi$ values are
forbidden, while obviously for $r\geq R+L/2$ all orientations are allowed.

\subsection{Inside a cylinder}

In contrast to the previous case, here the accessible $\phi$ range  is
centered around the cylinder axis, i.e., it has the form
$I_\phi=[\phi_c,\pi-\phi_c]\cup[\pi+\phi_c,2\pi-\phi_c]$. Since
tangential contact is not possible, the classification is a
little bit simpler. In the region of interest $|R|-L/2\leq r\leq |R|$ we obtain:
\begin{itemize}
\item[A,] for $\cos\theta\geq 2(r-|R|)/L$: $\cos\phi_c=\pi/2$,
\item[B,] for $r\geq r_c$ and $\cos\theta\leq 2(r-|R|)/L$ or for $r\leq r_c$ and
$(R^2-L^2/4-r^2)/(L r)\leq \cos\theta\leq 2(r-|R|)/L$: 
\begin{equation}
  \cos\phi_c=\frac{2\sqrt{R^2-(r+L/2\cos\theta)^2}}{L\sin\theta},
\end{equation}
\item[C,] for $r\leq r_c$ and $\cos\theta\leq (R^2-L^2/4-r^2)/(L r)$:
$\cos\phi_c=0$.
\end{itemize}

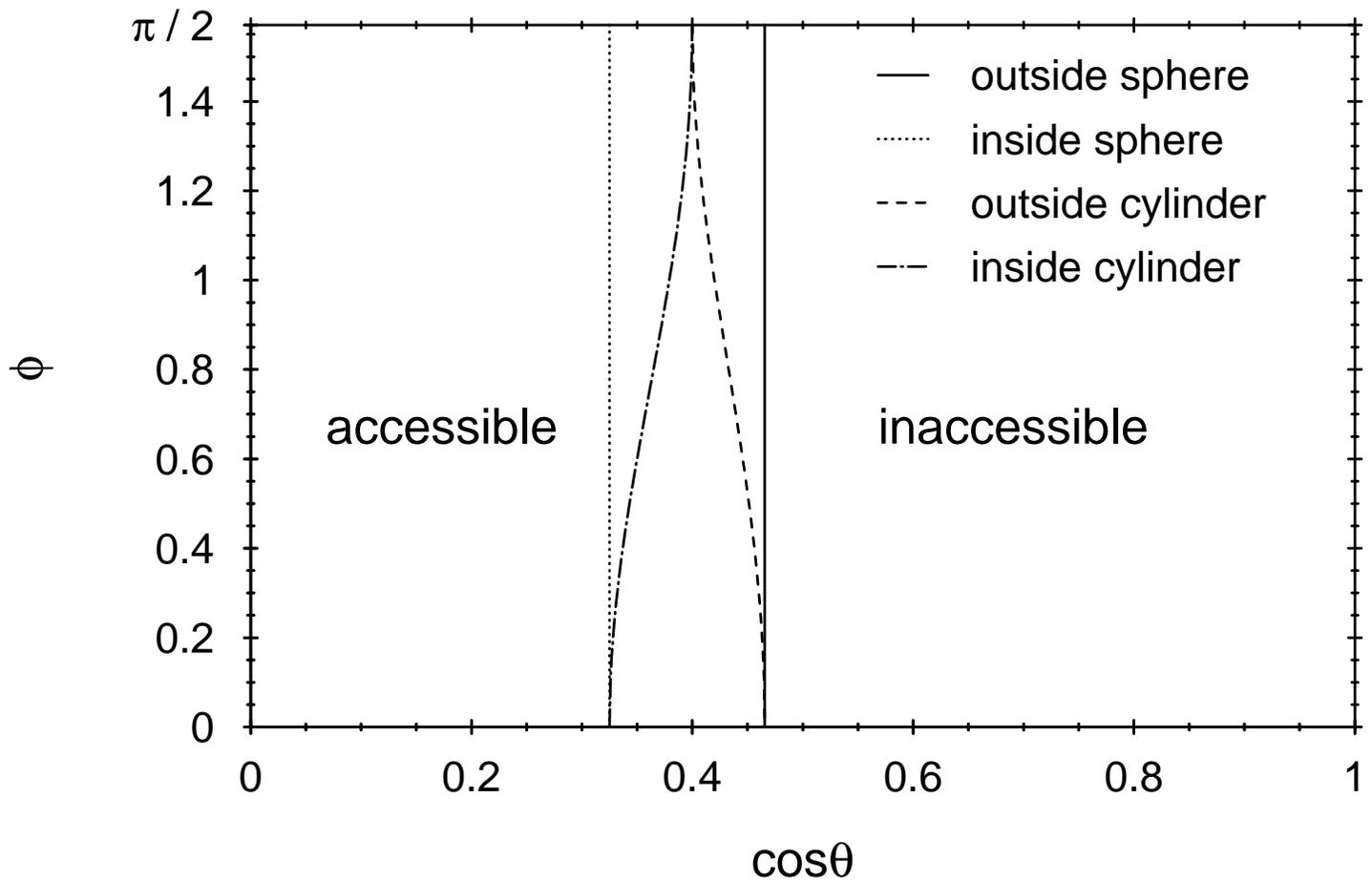
\begin{figure}
\caption{Accessible orientational space for a rod near walls of
different geometries at a fixed distance $z/L=0.2$ and radius
$|R|/L=3$. Spatial angles to the left of the lines are allowed, those to
the right are forbidden by the presence of the wall. There is no
dependence on the azimuthal angle $\phi$ for spherical walls. For a
cylinder at $\phi=0$ ($\phi=\pi/2$) the wall is effectively spherical
(planar) which explains the common end points of the various lines.}
\label{fig:rhoidsc}
\end{figure}

\bigskip
In \abbref{fig:rhoidsc} we compare the accessible orientational space
in these four cases for the same radius $|R|/L=3$ and the same distance from the
surface $z/L=0.2$. Naturally this space is largest outside a sphere
and smallest inside a sphere. One also notices that for a cylinder the
$\phi$ dependence is actually restricted to a rather small range of
values for
$\cos\theta$ while for most polar angles $\theta$ either none or all
azimuthal angles are allowed. We emphasize again that the forbidden
regions are also strictly forbidden for interacting rods whose
profile is no longer constant within the allowed region.

\end{appendix}

%%%%%%%%%%%%%%%%%%%%%%%%%%%%%%%%%%%%%%%%%%%%%%%%%%%%%%%%%%%%%%%%%%%%%%
%%%%%%%%%%%%%%%%%%%%%%%%%%%%%%%%%%%%%%%%%%%%%%%%%%%%%%%%%%%%%%%%%%%%%%
%\bibliographystyle{prsty}
%\bibliography{rods}

%%%%%%%%%%%%%%%%%%%%%%%%%%%%%%%%%%%%%%%%%%%%%%%%%%%%%%%%%%%%%%%%%%%%%%
% FIGURES
%%%%%%%%%%%%%%%%%%%%%%%%%%%%%%%%%%%%%%%%%%%%%%%%%%%%%%%%%%%%%%%%%%%%%%

\end{document}